\documentclass[12pt,letterpaper]{article}
\usepackage[utf8]{inputenc}
\usepackage{amsmath}
\usepackage{mathptmx, courier,textcomp}
\usepackage{amsfonts}
\usepackage{amssymb}
\usepackage{graphicx}
\usepackage{cite}
\usepackage{ulem}
\usepackage{caption}
\usepackage{epstopdf}
\usepackage[top=1in,bottom=1in,left=1in,right=1in]{geometry} 
\usepackage{float}
\usepackage{hyperref}
\usepackage{booktabs}
\usepackage[export]{adjustbox}
\usepackage{float}
\usepackage{multirow}
\usepackage{pifont}
\usepackage[ ]{authblk}
\usepackage{chemarrow}
\usepackage{comment}
\usepackage{makecell}
\usepackage{physics}
\usepackage{xcolor}
\UseRawInputEncoding
\hypersetup{
    colorlinks=true,
    linkcolor=blue,
    filecolor=magenta,      
    urlcolor=blue,
    pdftitle={Overleaf Example},
    pdfpagemode=FullScreen,
} 
\begin{document}

%
%

\title{Mechanistic Insights into Enhanced Alkaline Oxygen Evolution on Zn-Al Alloy Electrodes}
%
%

%
%
\author[1,$\dag$]{Abdul Ahad Mamun}
\author[1,$\dag$]{Rokon Uddin Mahmud}
\author[2]{Shahin Aziz}
\author[2]{Muhammad Shahriar Bashar}
\author[3]{Ahmed Sharif}
\author[*,1]{Muhammad Anisuzzaman Talukder}
\affil[1]{\small{Department of Electrical and Electronic Engineering \\

Bangladesh University of Engineering and Technology, Dhaka 1205, Bangladesh}} 
\affil[2]{\small{Institute of Energy Research and Development (IERD) \\

Bangladesh Council of Scientific and Industrial Research (BCSIR), Dhaka, 1205, Bangladesh}} 
\affil[3]{\small{Department of Materials and Metallurgical Engineering \\

Bangladesh University of Engineering and Technology, Dhaka 1000, Bangladesh}}

\affil[$\dag$]{\small{\it{These authors contributed equally to this work.}}}
\affil[*]{\small{\it{anis@eee.buet.ac.bd}}}
%
%

\date{ }
\maketitle
\sloppy

%
%
\begin{abstract}

Electrochemical water electrolysis, which produces clean energy carriers to mitigate carbon emissions, lacks suitable, low-cost electrodes for efficient oxygen evolution reaction (OER) in alkaline water splitting. To address this challenge, we developed zinc-aluminum (Zn-Al) alloy electrodes with varying Al contents up to 20 wt.\% via powder metallurgy method and conducted electrochemical measurements of the OER in alkaline solution to investigate their catalytic performance. We also performed first-principles calculations to examine their thermodynamic phase stability and electronic structures. Both theoretical and experimental results indicated that incorporating $\geq 20$ wt.\% Al into Zn led to thermodynamic phase instability and secondary-phase segregation in Al-rich regions, limiting reaction kinetics and reducing catalytic efficiency. Although the Al content of 5 wt.\% into Zn (Zn$_{0.95}$Al$_{0.05}$)  exhibited favorable thermodynamic and electronic characteristics, but its electrochemical performance was inefficient and poor due to inadequate reaction active sites on the surface. In contrast, the 10 wt.\% (Zn$_{0.9}$Al$_{0.1}$) and 15 wt.\% Al into Zn (Zn$_{0.85}$Al$_{0.15}$) showed approximately three- and two-fold increases in anodic exchange current density ($J_{0,a}$) relative to pure Zn, respectively. Additionally, the anodic overpotential losses ($\eta_{0,a}$) measured at a current density ($J$) of 12 mAcm$^{-2}$ were 0.240 V for Zn$_{0.9}$Al$_{0.1}$ and 0.5603 V for Zn$_{0.85}$Al$_{0.15}$, significantly lower than that of pure Zn ($\eta_{0,a} = 1.086$ V). While Zn$_{0.9}$Al$_{0.1}$ and Zn$_{0.85}$Al$_{0.15}$ showed similar charge transfer resistance ($R_{\rm CT}$), Zn$_{0.9}$Al$_{0.1}$ demonstrated superior reaction kinetics and lower overpotential losses across all samples tested. Furthermore, the improved kinetics and reduced overpotential of the Zn-Al alloys favorably compare with those of other transition-metal-based catalysts, including Fe-Co-Ni-Mo alloys and Fe-doped CuO. 


\end{abstract}
%
%

%
%
\section{Introduction}

The urgent transition from conventional fossil fuels to zero-carbon chemical fuels is essential to mitigate carbon emissions, which pose severe threats to human life and nature, as emphasized at the UN climate change conference (COP30) in Brazil \cite{fearnside2025cop,pitsch2024transition,mamun2024techno}. Among the various zero-emission energy carriers, hydrogen (H$_2$) fuel stands out as one of the most promising options due to its higher gravimetric energy density, availability from abundant resources, and its inherently environmentally friendly nature \cite{mamun2025advancing,mamun2024enhancing}. Currently, approximately 96\% of H$_2$ is produced from fossil fuels, such as methane and coal, with only about 4\% through direct water electrolysis \cite{segovia2025green}. Therefore, increasing the proportion of H$_2$ production through water splitting has become a primary goal to reduce dependence on fossil fuels and diminish environmental pollution. Additionally, H$_2$ production via electrochemical processes plays a crucial role in energy conversion and storage, helping mitigate the intermittency associated with renewable energy sources like wind and solar \cite{qian2024recent,dutta2018designing}. However, cost-effective, efficient, and durable water electrolysis techniques face significant challenges, including sluggish reaction kinetics and reduced durability of electrode materials \cite{rahman2025enhanced,ruan2025technologies}.

Electrodes are essential components in water electrolysis and serve as the active surfaces for evolution reactions that convert electrical energy into chemical fuels \cite{rahman2025enhanced,wan2023key}. The efficiency of charge transfer and the chemical transformations of reactive species are influenced by the electrode's characteristics, including their electronic properties, surface area, chemical composition, and catalytic activity \cite{wan2023key}. In water electrolysis, the cathode facilitates the hydrogen evolution reaction (HER), and the anode drives the oxygen evolution reaction (OER) \cite{mamun2025advancing,mamun2023effects}. The HER involves a two-electron transfer process to produce H$_2$, resulting in relatively fast reaction kinetics. In contrast, the OER involves multiple reaction intermediates and requires the transfer of four electrons to complete the reaction, resulting in slower reaction kinetics. This reduced rate of the OER compared to the HER makes it the rate-determining step in the overall water-splitting processes, causing larger overpotential losses ($\eta_{0}$) for producing H$_2$ and oxygen (O$_2$) gases \cite{exner2019beyond}. Additionally, the poor electrical conductivity and degradation of the electrode materials further limit the overall efficiency of water splitting, posing significant challenges for large-scale H$_2$ production \cite{yao2021strategy,hussain2025physics}. To overcome these challenges, it is essential to develop highly active, stable, and low-cost electrodes that accelerate reaction kinetics and improve the overall efficiency of water-splitting systems.

In recent decades, noble metal-based catalysts, such as ruthenium (Ru) and iridium (Ir), as well as their oxide forms (RuO$_2$ and IrO$_2$), have demonstrated excellent properties for adsorbing oxygenated intermediates and exhibiting favorable surface redox characteristics during the OER process \cite{shi2019robust,qin2024ru}. These advancements allow for rapid reaction kinetics and help minimize $\eta_{0}$. Additionally, their high intrinsic catalytic activity and remarkable resistance to oxidative degradation make them standard catalysts for water electrolyzer systems. However, the limited availability and high cost of these materials present significant challenges to their widespread use in large-scale H$_2$ production \cite{yan2016review,datye2021opportunities}. 

Recently, nanostructured carbon materials have been explored for OER electrocatalysis, as they serve as conductive supports and offer high surface area and exceptional conductivity \cite{asefa2021nanostructured}. Despite these advantages, the high anodic potential for the OER process can lead to oxidative degradation in metal-free carbon frameworks, ultimately limiting their catalytic activity and chemical stability. Conversely, transition metals (TMs) show promise as alternatives to noble metals due to their lower cost, natural abundance, diverse morphologies, and tunable redox properties \cite{roy2024engineered,das2022transition}. However, TMs generally exhibit low intrinsic OER activity due to their narrow d-bands and polaronic charge transport \cite{li2025electron}. These phenomena result in increased charge-transfer resistance ($R_{\rm CT}$) and suboptimal adsorption energetics for intermediate species. Therefore, it is essential to apply cost-effective and efficient engineering techniques to overcome the limitations of TMs.

Several transition metal-based electrocatalysts, including copper (Cu), cobalt (Co), nickel (Ni), zinc (Zn), and iron (Fe), have been used for the OER activity owing to their unique electronic structures and dynamic surface transformations \cite{noor2021recent}. Edao et al.~developed heterostructured electrocatalysts by combining Ni-Fe layered double hydroxides with nickel sulfide, achieving a low $\eta_0$ of $0.220$ V at a current density ($J$) of $100$ mA cm$^{-2}$ \cite{edao2024nickel}. The introduction of erbium (Er) into Ni-Fe layered double hydroxides led to improved OER performance, driven by changes in the d-bands, superior defect chemistry, and increased active sites \cite{zhu2023improving}. On the other hand, transition-metal-based alloy electrodes, e.g., Ni-Fe and Fe-Co, exhibit improved electrocatalytic activity and high efficiency for the OER due to their tunable 3d electronic structures, which optimize the adsorption energies for oxygenated intermediates \cite{lim2020bimetallic, zhu2020porous}. 

Beyond the Fe and Ni materials, spinel Zn$_{1-x}$Co$_{x}$Fe$_{2}$O$_{4}$ and ZnCo$_{2}$O$_{4}$-derived electrodes have been recently investigated for water splitting, demonstrating that the incorporation of Zn with Co and Fe can effectively tune the electronic structure and enhance the active sites for the OER \cite{shahzadi2025zn}. Thus, Zn-based electrodes are rapidly gaining recognition in water electrolysis, thanks to their abundance, high theoretical electrochemical capacity, tunable surface chemistry, and low-cost manufacturing process \cite{singh2025zinc}. These Zn-based electrodes can also form oxide and hydroxide layers that undergo self-adaptive surface oxidation, which modulates the concentration of local species and intermediate adsorption within the diffusion layer \cite{hao2020deeply,singh2025recent}. However, the intricate relationships among evolving OER activity, charge transfer kinetics, and the structural properties of modified Zn-based electrodes remain largely unexplored. This aspect highlights the need for systematic investigations and a deeper understanding of electrochemical roles and transformation pathways in emerging water-splitting processes.

%
In this study, we developed Zn-based binary alloy electrodes incorporating aluminum (Al) in varying compositions for alkaline water electrolysis, combined with first-principles calculations and experimental methods. Al was chosen as a secondary element for Zn-Al alloys due to its favorable thermodynamic properties and its ability to modify electronic structure and surface chemistry, thereby influencing charge transfer and reaction energetics during the OER process. Firstly, we conducted density functional theory (DFT) simulations to identify the optimal Al composition by evaluating thermodynamic phase stability parameters, including formation enthalpy ($E_f$) and cohesive energy ($E_{\text{coh}}$). To investigate electronic properties, we also calculated the electronic density of states (DOS), work function ($\phi$), and the atom's ion charge. Our theoretical analysis revealed that Zn with up to $\sim$15 wt.\% Al demonstrated thermodynamic phase stability, whereas a composition of $\gtrsim 20$ wt.\% Al showed phase instability. Next, we synthesized Zn-Al alloys with Al compositions varying up to 20 wt.\% using the powder metallurgy method. To characterize the materials, we employed X-ray diffraction (XRD), optical microscopy, and scanning electron microscopy (SEM). We also conducted electrochemical studies on the developed Zn-Al alloy electrodes utilizing cyclic voltammetry (CV), polarization curves (PC), Tafel analysis, electrochemical impedance spectroscopy (EIS), and chronoamperometry (CA). Finally, this study provided integrated theoretical and experimental insights into the mechanisms enabling enhanced charge-transfer kinetics and catalytic activity of Zn-Al alloy electrodes for OER.

Our results revealed that Zn-Al alloys with 10 wt.\% (Zn$_{0.9}$Al$_{0.1}$) and 15 wt.\% Al (Zn$_{0.85}$Al$_{0.15}$) exhibited more than twice the anodic exchange current density ($J_{0,a}$) compared to pure Zn. The $R_{\rm CT}$ for Zn$_{0.9}$Al$_{0.1}$ and Zn$_{0.85}$Al$_{0.15}$ significantly reduced to 1.85 and 1.76 $\Omega$cm$^2$, respectively, much lower than that of 9.52 $\Omega$cm$^2$ for pure Zn. Additionally, the anodic overpotential losses ($\eta_{0,a}$) for Zn$_{0.9}$Al$_{0.1}$ and Zn$_{0.85}$Al$_{0.15}$ in a 1 M KOH solution were $0.240$ V and $0.5603$ V at $J=12$ mAcm$^{-2}$, respectively, significantly lower than $1.086$ V observed for pure Zn. These improvements suggest that Zn-Al alloy electrodes outperform pure Zn and are comparable with other complex TM-based catalysts, such as Fe-Co-Ni-Mo alloys, Ni-Fe alloys, and Fe-doped CuO. Overall, these results indicate that Zn-Al alloy electrodes are promising, low-cost, and efficient materials with accelerated reaction kinetics for large-scale water-splitting applications. 


%
%
\section{Methodology}
\subsection{Computational Approach}

We conducted all first-principles calculations using a self-consistent ab initio method within the framework of spin-polarized DFT, implemented in the Quantum Espresso (QE) code  \cite{giannozzi2009quantum}. The exchange-correlation interactions were calculated using the generalized gradient approximation (GGA) approach with the Perdew-Burke-Ernzerhof (PBE) functional \cite{giannozzi2009quantum,giannozzi2017advanced}. The projected augmented wave (PAW) method was utilized to determine core electrons, while valence electrons were estimated using Kohn-Sham (KS) single-electron orbitals, which were solved by expanding a plane-wave basis with a cut-off kinetic energy of 70 Ry \cite{kresse1999ultrasoft}. To resolve fine details of electron distributions, the charge density cut-off was considered at twelve times the wave function cut-off energy. To aid convergence and stabilization of electronic occupations, we applied the Marzari-Vanderbilt (MV) smearing method with a smearing width of 0.01 Ry. Brillouin-zone integration was performed using a generalized Monckhorst-Pack $k$-point mesh of $7 \times 7 \times 7$ for both geometry optimization and electronic structure calculations \cite{monkhorst1976special}. The DFT-D3 method developed by Grimme et al.~was employed to account for long-range van der Waals interactions \cite{grimme2010consistent}. We performed structural relaxations using the Broyden-Fletcher-Goldfarb-Shanno (BFGS) algorithm until the atomic forces remained below 0.001 eV/{\AA} and the change in total energy was less than 10$^{-7}$ eV.

%
\begin{figure}[hbt]
    \centering
    \includegraphics[width =0.65\linewidth]{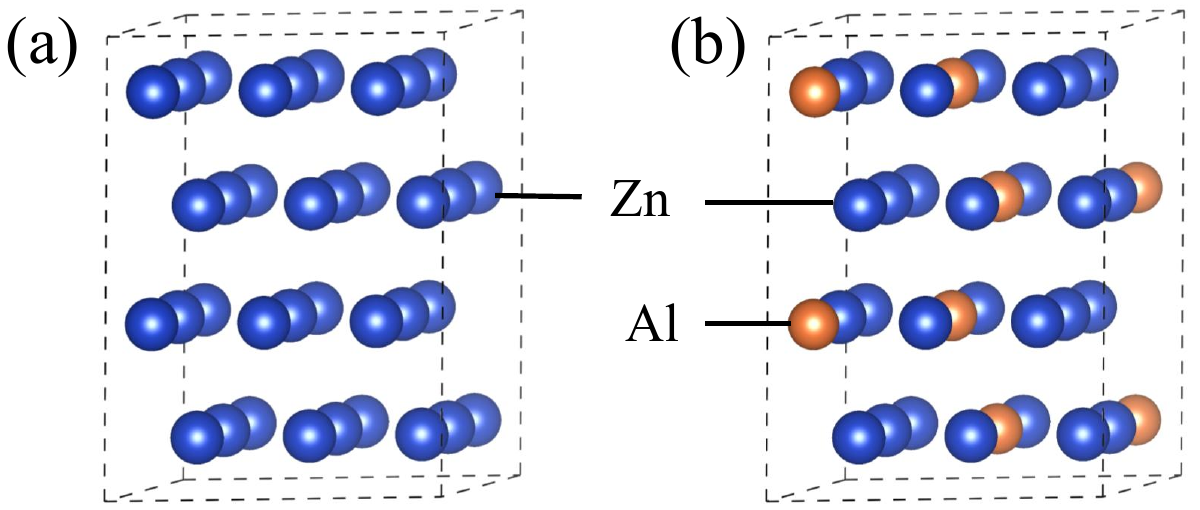}
    \caption{Hexagonal crystal structure of (a) pure Zn and (b) Zn-Al alloy electrodes.}
    \label{Fig01}
\end{figure}
%

We designed the alloy electrodes using a hexagonal crystal structure of pure Zn (space group P6$_3$/mmc [194]) and a cubic crystal structure of Al (space group Fm$\bar{3}$m [225]). Figure \ref{Fig01}(a) illustrates the hexagonal close-packed supercell of pure Zn, while Fig.~\ref{Fig01}(b) displays the supercell of Zn-Al alloy structures incorporating both Zn and Al atoms. We created a Zn supercell matrix with Al compositions of 5, 10, 15, and 20 wt.\% (refer to Fig.~S1), and the required number of Al atoms can be calculated using 
%
\begin{equation}
x ({\rm wt.}\%) = \frac{N_{\rm Al} \times M_{\rm Al}}{(N - N_{\rm Al}) \times M_{\rm Zn} + N_{\rm Al} \times M_{\rm Al}} \times 100,
\end{equation}
%
where $N$ and $N_{\rm Al}$ are the total number of atoms and the number of Al atoms within the supercell, respectively, and $M_{\rm Zn}$ and $M_{\rm Al}$ are the atomic masses of Zn and Al, respectively.

The incorporation of Al into Zn can alter the electronic structure, local atomic coordination, and overall energetics of the alloys, thereby directly affecting the structural integrity and catalytic behavior of the materials. To systematically understand these effects, it is crucial to investigate the thermodynamic and structural properties of the alloys at the atomic level. The $E_f$ and $E_{\rm coh}$ serve as essential parameters for assessing the thermodynamic stability of the alloys. $E_f$ quantifies the energetic driving force for alloying, which indicates the phase stability and structural integrity of the alloy compared to its pure element \cite{qi2023effect}. On the other hand, $E_{\rm coh}$  reflects the strength of atomic bonding within the alloy lattice, which is related to the material's mechanical durability and resistance to degradation in electrochemical environments \cite{yaghmaee2017thermodynamics}. We calculated both $E_f$ and $E_{\rm coh}$ for the designed Zn-Al alloy electrodes as given by \cite{lu2021first}
%
\begin{subequations}
\begin{align}
  & E_{f} = \frac{1}{N} \left[ E_{\rm total} - N_{\rm Zn} \times E_{\rm Zn}^{\rm solid} - N_{\rm Al} \times E_{\rm Al}^{\rm solid} \right], \\
  & E_{\rm coh} = \frac{1}{N} \left[ E_{\rm total} - N_{\rm Zn} \times E_{\rm Zn}^{\rm atom} - N_{\rm Al} \times E_{\rm Al}^{\rm atom} \right] ,
\end{align}
\end{subequations}
%
where $E_{\rm total}$ is the total energy of the alloy supercell, and $E_{\rm Zn}^{\rm solid}$ and $E_{\rm Al}^{\rm solid}$ are the solid phase energies of Zn and Al atoms, respectively. $N_{\rm Zn}$ denotes the total number of Zn atoms within the alloy supercell. Additionally, $E_{\rm Zn}^{\rm atom}$ and $E_{\rm Al}^{\rm atom}$ are the energies of isolated Zn and Al atoms, respectively.

We also computed the electronic DOS, which provides a quantitative understanding of band dispersion and the density of carrier states near the Fermi level \cite{rahman2025enhanced}. This insight allows us to assess metallic properties, electrical conductivity, and the availability of energy states that influence adsorption processes and reaction kinetics. The parameter $\phi$ was derived from the electrostatic potential and the position of the Fermi level. This parameter reflects the strength of the surface electric field and the thermodynamic driving force for interfacial charge transfer in catalytic processes \cite{chen2024work}. Additionally, we calculated the atom's ion charges using Bader charge analysis, which represents charge density at the surface and correlates with bond polarization and the electronegativity of the active site.

%
\begin{figure}[hbt]
    \centering
    \includegraphics[width =1.0\linewidth]{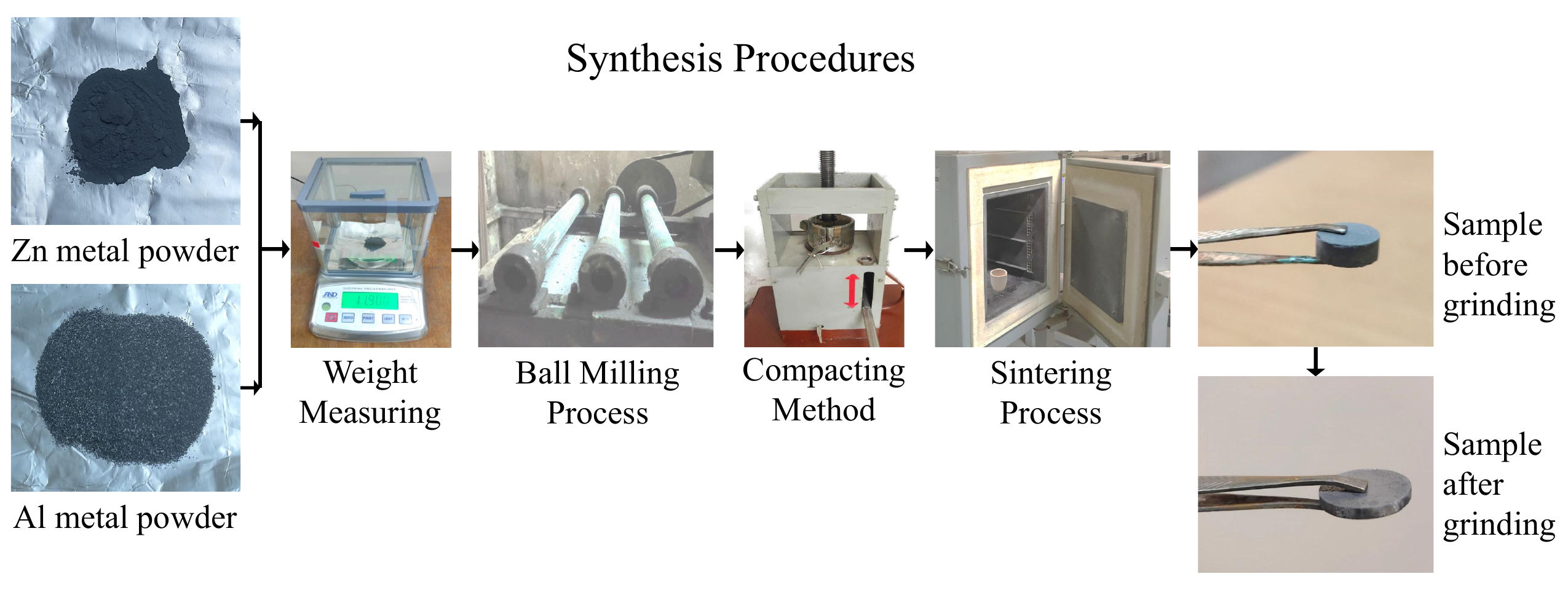}
    \caption{Schematic illustration of the synthesis process for Zn-Al alloy electrodes.}
    \label{Fig02}
\end{figure}
%
%

%
\subsection{Experimental Methods}

\subsubsection{Materials Preparation}

The Zn-Al alloys were fabricated using a powder metallurgy technique that involves blending metal powders, compacting them into electrode shapes, and performing a heat treatment. Initially, Zn and Al metal powders (Sigma-Aldrich, $\ge 99.5$\% purity) were accurately weighed according to the specified composition ratios. To achieve a homogeneous distribution and refine the metal particles, the mixed powders underwent mechanical ball milling for six hours using contaminant-free zirconium balls, with a ball-to-powder mass ratio of at least 6:1. Next, the milled alloy powders were compacted into circular disk shapes using a hydraulic compression machine operating at 300 atm during six minutes. This uniaxial compaction increased the particle packing density and facilitated mechanical stability. Each circular disk had a radius of 15 mm and a thickness of 2.5 mm. 

The compacted disks were subsequently sintered at 370 $^\circ$C for 1 hour, wrapped in graphite foil and supported within a closed ceramic setter. This setup minimized oxidation and maintained a controlled micro-environment. The thermally activated sintering process enables diffusion bonding, pore shrinkage, and metallurgical bonding between adjacent particles, thereby increasing densification, enhancing structural integrity, and improving mechanical strength. Finally, the sintered disks underwent a sequential surface preparation procedure, and the electrode surfaces were polished with emery paper grit sizes 300, 600, 800, 1000, 1200, and 1500 to achieve a smooth surface with minimal defects. The prepared surfaces were then cleaned by rinsing with deionized water, acetone, and ethanol to remove residual particles and inorganic contaminants, ensuring that the Zn-Al alloy electrodes were highly clean and suitable for electrochemical testing. Figure \ref{Fig02} provides a detailed schematic representation of the synthesis process for producing the Zn-Al alloy electrodes.

\subsubsection{Materials Characterization}

The characterization of materials provides valuable insights into the relationships between micro-structure, phase transformations, defect generation, and local surface chemistry. To assess the crystallographic structure and lattice parameters of the synthesized alloys, we conducted XRD analysis using an ARL EQUINOX 1000 X-ray diffractometer. This technique included a sealed Cu anode tube operating at 45 kV and 40 mA, producing Cu K$\alpha$ radiation with a wavelength of $\lambda = 1.54060$ {\AA}. The Bragg-Brentano $\theta$--$2\theta$ geometry, along with a high-resolution goniometer and a position-sensitive detector (PSD), was utilized to ensure high-quality data collection with minimized instrumental broadening. The diffraction patterns were accumulated over a scanning range of $ 20^\circ \le 2\theta \le 100^\circ$, with a step size of 0.01$^\circ$, while optimizing the dwell time to achieve a high signal-to-noise ratio. We subsequently calculated the crystallite size and lattice microstrain using the Scherrer approximation and Williamson-Hall methods.

To obtain an initial visualization of the macro-structure, we performed optical microscopy under bright-field illumination, which enabled us to examine grain morphology, surface segregation, and deformation structures in the synthesized alloys. Optical microscopy is particularly useful for distinguishing metallurgical phases and identifying macro-structural abnormalities based on optical reflectivity and contrast mechanisms. However, it is limited to micrometer-scale spatial resolution. To capture finer structural and morphological details beyond the optical diffraction limit, we conducted SEM analysis using a Hitachi S-3400N Variable Pressure SEM operated at 25 kV, considering both secondary electron (SE) and back-scattered electron (BSE) imaging modes. This method allowed us to investigate various microstructural features, including nanoscale grain morphology, porosity distribution, dendritic and inter-metallic phase formation, and localized deformation textures. Overall, these integrated approaches provide a comprehensive multiscale assessment of structure-property relationships and a deeper understanding of the fundamental mechanisms governing phase evolution, micro-structural refinement, and overall material performance.

\subsubsection{Electrochemical Studies}

Electrochemical analyses are used to assess the catalytic activity, reaction kinetics, and durability of alloy electrodes. We performed electrochemical measurements using a Corrtest electrochemical workstation (CS350M EIS potentiostat) at a temperature ($T$) of 25$^\circ$ and a pressure ($P$) of 1 atm in a 1 M KOH solution. The experimental setup used a three-electrode configuration within a standard 100 ml H-cell. In this setup, the alloy electrodes acted as the working electrode (WE), a platinum (Pt) plate served as the counter electrode (CE), and a saturated calomel electrode (SCE) was employed as the reference electrode (RE). The electrochemical testing involved various techniques, including CV, linear sweep voltammetry (LSV), potentiodynamic polarization resistance (LPR), and EIS, as well as assessments of chemical durability and stability. We started these analyses after the stabilization of the open-circuit potential (OCP) within the electrochemical system.

CV was conducted within a potential window of $-2.5$ to $+0$ V vs. SCE, with a scan rate of $5$ mV/s. The LSV curves were recorded with the range of $-0.8$ to $+1.5$ V vs. SCE at a scan rate of $20$ mV/s, and the LPR was performed from $-2.5$ to $+1.2$ V vs. SCE, operating at a scan rate of $10$ mV/s. We conducted EIS analysis over a frequency range of $10000$ to $0.05$ Hz, with an amplitude of $5$ mV. The chemical stability of alloy electrodes was evaluated through chronoamperometry measurements at an applied potential ($V_{\rm app}$) of $+0.8$ V vs. SCE.

The anodic Tafel slope ($b_a$) was obtained from the LPR curve. To calculate the charge transfer coefficient ($\alpha_n$) and $J_{0,a}$, we used the following equations \cite{neyerlin2007study,hussain2025physics}
%
\begin{subequations}
\begin{align}
  & \alpha_n = \frac{RT}{b_a nF}, \\
  & \eta_{0,a} = -\frac{RT}{\alpha_{n} \, nF} \ln(J_{0,a}) + b_a \ln(J),
\end{align}
\end{subequations}
%
where $R$, $F$, and $n$ represent the universal gas constant, the Faraday constant, and the number of electrons transferred, respectively. The polarization resistance ($R_P$) was determined from the LPR curve over a potential range of $\pm 200$ mV relative to the corrosion potential ($E_{\rm corr}$). The $R_{\rm CT}$ was obtained from EIS data using the Randles circuit model, fitted in CS Studio Analysis software.

%
%
\section{Simulation Results}

\subsection{Thermodynamic Phase Stability}

The accurate estimation of lattice parameters is required to precisely define interatomic forces and distances within a crystal structure, enabling proper measurement of electronic and thermodynamic properties. In our study, we conducted a comprehensive analysis of the lattice parameters for primitive Zn and Al cells using the GGA-PBE method to determine the precise crystal structures and validate the accuracy of our computational approaches. For the primitive Zn cell, the simulated lattice parameters were $a = b = 2.647$ {\AA} and $c = 4.961$ {\AA}, which closely align with the experimental values of $a = b = 2.67$ {\AA} and $c = 4.95$ {\AA} \cite{nuss2010structural}. Additionally, the calculated $c/a$ ratio of $1.87$ nearly matched the experimental ratio of $1.86$, demonstrating consistency in our calculations \cite{wedig2007structural}. Similarly, the obtained lattice parameter of the primitive Al cell was $a = b = c = 4.03$ {\AA}, while the experimental value was $a = b = c = 4.05$ {\AA} \cite{tang2009surface}. Notably, the differences between the computed and experimentally observed values were below 1\% for both Zn and Al, confirming the validity and reliability of our computational methodology.

%
%
\begin{table}[hbt]
\centering
\caption{Formation energy ($E_f$), cohesive energy ($E_{\rm coh}$), work function ($\phi$), and Bader charge analyses of the designed Zn-Al alloy electrodes with varying Al contents.}
\resizebox{0.75\textwidth}{!}{%
\begin{tabular}{c c c c c c}
\Xhline{3\arrayrulewidth}
    Electrode Name & $E_f$ & $E_{\rm coh}$ & $\phi$ & \multicolumn{2}{c}{Bader charge ($e$C)}  \\
    \cline{5-6}
      &  (eV) & (eV/atom) & (eV) & Zn & Al \\
    \Xhline{2\arrayrulewidth}
    Pure Zn & -- & $-1.0738$ &  $4.280$ & $-0.000088$ & -- \\
    Zn$_{0.95}$Al$_{0.05}$ & $-0.2765$ & $-1.0125$ & $4.082$ & $-0.091124$ & $+0.637657$  \\
    Zn$_{0.90}$Al$_{0.10}$ & $-0.2502$ & $-0.9431$ & $4.254$ & $-0.165546$ & $+0.496493$ \\
    Zn$_{0.85}$Al$_{0.15}$ & $-0.1596$ & $-0.9148$ & $4.489$ & $-0.111814$ & $+0.246094$ \\
    Zn$_{0.80}$Al$_{0.20}$ & $+0.0158$ & $-0.8724$ & $4.172$ & $-0.155384$ & $+0.259178$ \\
    \Xhline{3\arrayrulewidth}
\end{tabular}}
\label{Table1}
\end{table}
%
%

The incorporation of Al into Zn to form Zn-Al alloys resulted in a slight increase in the lattice parameters compared to pure Zn. This expansion occurs mainly due to lattice distortions and changes in local interatomic distances induced by the substitution of Zn atoms with Al atoms \cite{kobayashi2018lattice}. However,  the Zn$_{0.8}$Al$_{0.2}$ alloy, which contains 20 wt.\% Al, exhibited slightly decreased lattice parameters, likely indicating thermodynamic destabilization. This behavior suggests that the solid solubility limit of Al in the Zn matrix may have been exceeded at higher Al concentrations, leading to elevated lattice strain and potential phase segregation. To investigate the phase stability further, we calculated $E_f$ for all compositions of Zn-Al alloys, as illustrated in Table \ref{Table1}. We found that Zn$_{0.8}$Al$_{0.2}$ had a positive $E_f$ value of $+0.0158$ eV, indicating thermodynamic phase instability and a tendency to form a metastable alloy. In contrast, the negative $E_f$ values for Zn$_{0.95}$Al$_{0.05}$, Zn$_{0.9}$Al$_{0.1}$, and Zn$_{0.85}$Al$_{0.15}$ confirm strong thermodynamic phase stability, with Zn$_{0.95}$Al$_{0.05}$ showing the lowest $E_f$ of $-0.2765$ eV. The transition of $E_f$ values from highly negative to less negative or positive with increasing Al composition can be attributed to higher elastic strain energy, repulsive chemical interactions, and the destabilization of the homogeneous solid solution \cite{chen2015phase}. 

To further understand the bonding characteristics within the crystal structure, we determined $E_{\rm coh}$ relative to the solid-phase system energy, as presented in Table \ref{Table1}. All Zn-Al alloys demonstrated negative $E_{\rm coh}$ values, indicating that external energy is required to separate the solid into isolated free atoms. These negative values also affirm strong interatomic bonding, which correlates with enhanced mechanical properties, such as hardness and elastic modulus. However, while Zn$_{0.8}$Al$_{0.2}$ displayed a negative $E_{\rm coh}$, its positive $E_f$ highlights thermodynamic instability, a tendency for phase transformations, and the presence of structural disorder. These factors may result in anomalous and inferior electronic and catalytic properties.

%
\begin{figure}[hbt]
    \centering
    \includegraphics[width =0.93\linewidth]{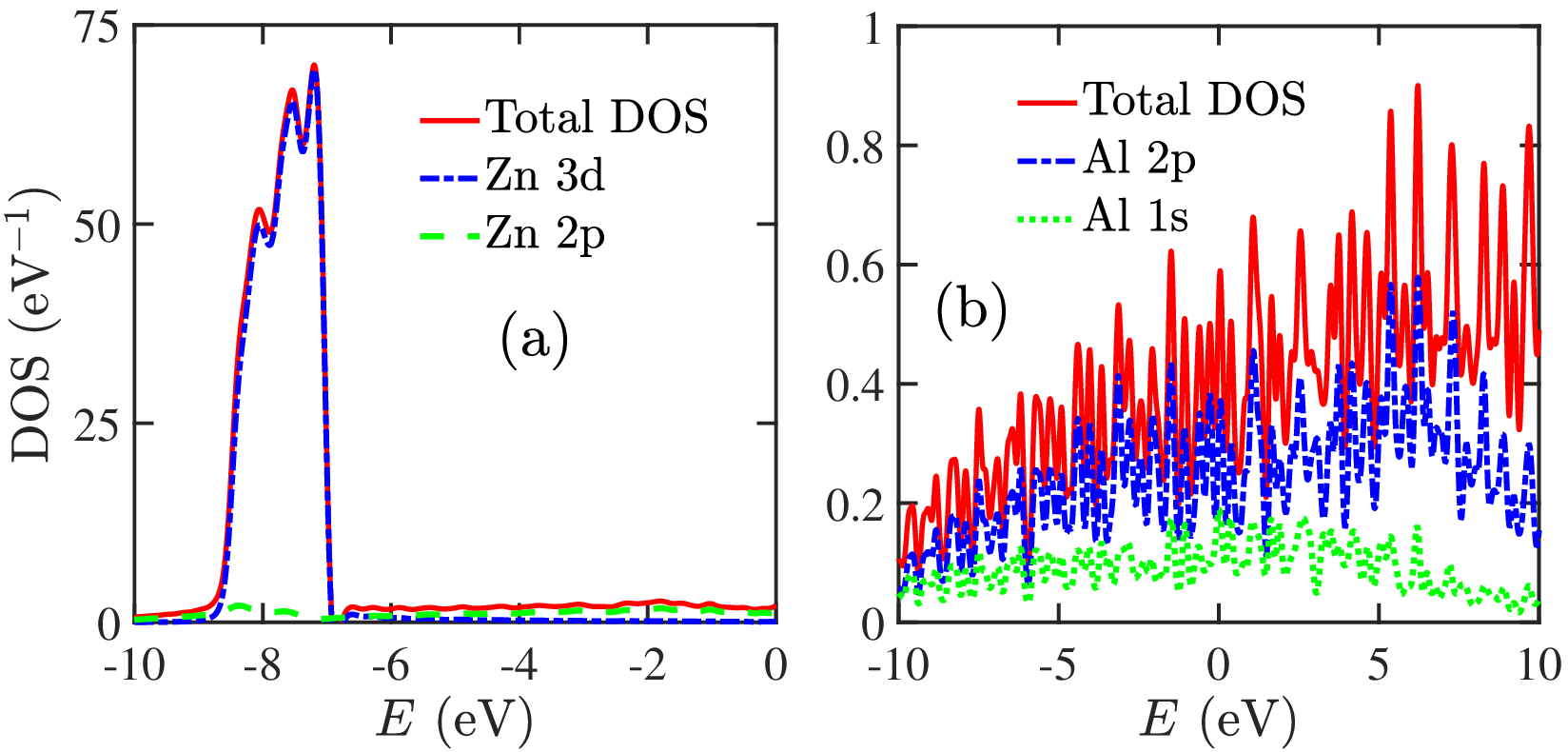}
    \caption{Electronic density of states (DOS) of (a) pure Zn using a $2 \times 2 \times 2$ supercell and (b) pure Al modeled with its unit cell crystal structure. For both of them, the Fermi level is set to zero.}
    \label{Fig03}
\end{figure}
%
%

%
\subsection{Electronic Properties}

We analyzed several important electronic properties, including the DOS, $\phi$, and atom's ion charge, to gain insights into electronic charge transport and catalytic activity. The DOS illustrates the distribution of electronic states and the degeneracy of charge carriers in the conduction band (CB), which helps qualitatively understand electrical conductivity and carrier concentration \cite{rahman2025enhanced,mamun2025improved}. A broader DOS across chemical energy with higher values indicates a larger number of conduction electrons, leading to improved electrical conductivity. Consequently, this enhancement increases $J$ with lower $V_{\rm app}$ while reducing ohmic losses. Figures \ref{Fig03}(a)--(b) show the DOS of pure Zn and Al, respectively. These electronic structures are consistent with previously reported literature, confirming the accuracy of our simulation method \cite{olguin2020total,jacobs2002bulk}. In the case of Zn, the d-orbitals dominate the total DOS, indicating that the conduction electrons mainly originate from d-orbitals. On the other hand, the p-orbitals primarily contribute to the total DOS for Al.

%
\begin{figure}[hbt]
    \centering
    \includegraphics[width =0.87\linewidth]{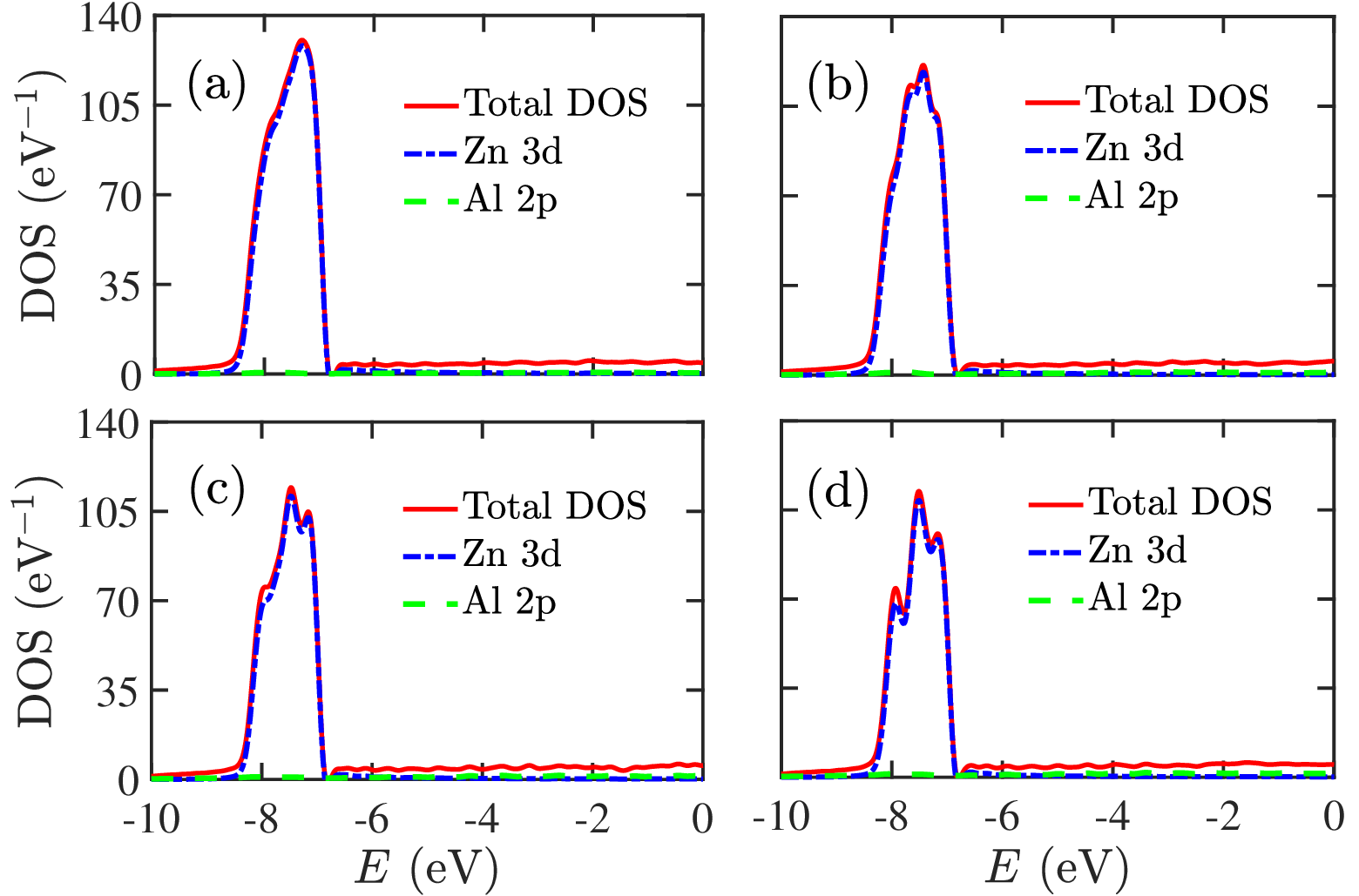}
    \caption{Electronic density of states (DOS) of (a) 5\% Al into Zn, (b) 10\% Al into Zn, (c) 15\% Al into Zn, and (d) 20\% Al into Zn electrodes using $2 \times 2 \times 2$ supercell crystal structure. For all of them, the Fermi level is set to zero.}
    \label{Fig04}
\end{figure}
%
%

Figure \ref{Fig04} shows the DOS and the projected DOS (PDOS) for Zn-Al alloys with Al compositions of 5, 10, 15, and 20 wt.\%. In all cases, the total DOS was predominantly influenced by the 3d orbitals of Zn, which highlights the significant role of these d-orbitals in carrier degeneracy. In contrast, the contribution of Al's 3p orbitals to the DOS was lower than that of Zn. Notably, the incorporation of Al into Zn broadened the DOS distribution across chemical energy levels and resulted in multiple distinct peaks at 10, 15, and 20 wt.\% Al compositions. This observation indicates a strong hybridization between Zn's 3d and Al's 3p states. Overall, these findings indicate the renormalization of the electronic structure driven by interactions between Zn and Al, resulting in superior charge-transport dynamics within the alloy materials.

The magnitudes of DOS for all Zn-Al alloys were approximately twice that of pure Zn, indicating stronger electron delocalization and a higher number of conduction states. This enhancement suggests that the alloy materials exhibited increased electrical conductivity. Among all the investigated compositions, the alloy with 5 wt.\% Al showed the highest DOS intensity. However, this intensity decreased slightly as Al content increased. This reduction may result from changes in the band structure, impurity scattering, and increased lattice disorder induced by Al.

The calculated $\phi$ and atomic ion charges offer a mechanistic understanding of catalytic behavior and charge transfer between the electrode and electrolyte. The adsorption and desorption processes of the evolution reaction can also be influenced by tuning $\phi$ of the electrodes \cite{zeradjanin2017balanced}. For the HER in alkaline media, an optimal $\phi$ enhances the accumulation of H$^+$ ions within the electric double layer (EDL) and promotes efficient proton-electron coupling, thereby improving catalytic activity. Conversely, for the OER in alkaline conditions, a relatively higher $\phi$ effectively stabilizes intermediates, such as $^*$OH, $^*$O, and $^*$OOH, which energetically favor the rate-determining steps of the reaction. The calculated $\phi$ for pure Zn was 4.28 eV, which closely aligns with literature values and falls within the optimal range for efficient adsorption and desorption processes \cite{heo2022modulation}. In all the designed compositions, the $\phi$ values slightly changed compared to pure Zn, as presented in Table \ref{Table1}. These minimal changes occur because the electronic structure of Zn predominantly influences the Fermi level, while Al contributes only minor perturbations to the electronic structure. These findings confirm that key intermediates effectively convert to other species, allowing for the completion of the evolution reactions without forming strong metal-oxygen bonds. 

In the atomic charge analysis, Zn atoms exhibited negative ionic charges, while Al atoms displayed positive ionic charges within the alloy materials, as shown in Table \ref{Table1}. This bipolar charge distribution creates a local built-in electric field that optimizes intermediate adsorption toward thermo-neutral values, thereby enhancing charge transfer between the electrode and electrolyte. Overall, these theoretical findings demonstrate that the alloy compositions outperform pure Zn in terms of electronic properties and catalytic activity.

\section{Experimental Results}

\subsection{Materials Physical Characterization}

\begin{figure}[hbt]
    \centering
    \includegraphics[width =0.87\linewidth]{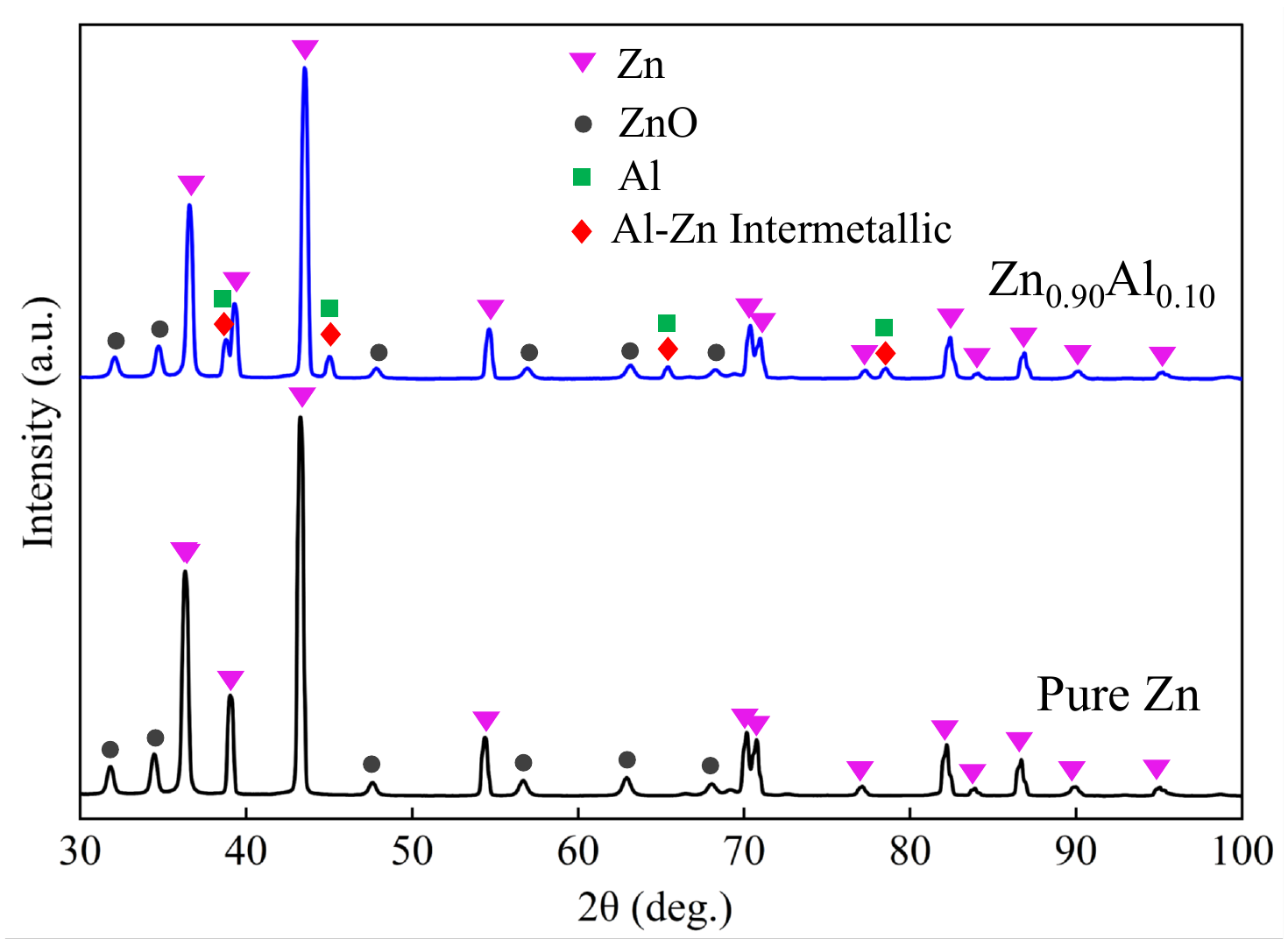}
    \caption{X-Ray Diffraction (XRD) pattern for pure Zn and 10 wt.\% Al into Zn alloy (Zn$_{0.9}$Al$_{0.1}$) electrodes.}
    \label{Fig05}
\end{figure}

The XRD pattern of the pure Zn sample displayed distinct diffraction features of a hexagonal close-packed (hcp) crystal structure, as illustrated in Fig.~\ref{Fig05}. The observed reflections correspond to the P6$_3$/mmc space group, which aligns with established crystallographic data \cite{masadeh2019total}. The prominent peaks at $2\theta$ $\approx 36.3^\circ$, $38.9^\circ$, $43.2^\circ$, $54.4^\circ$, $70.1^\circ$, and $82.2^\circ$ indicated the (002), (100), (101), (102), (103), and (112) planes of Zn, respectively. These peaks closely matched the standardized ICDD/PDF reference pattern (PDF 04-0831) with high accuracy \cite{nasui2025cold}. The narrow full-widths at half-maximum (FWHM) of these peaks suggested relatively large coherent diffraction domains, minimal microstrain, and a low defect density in metallic Zn. The calculated crystallite size of $20.54$ nm and microstrain of $3.8 \times 10^{-3}$ fell within the range of previously reported values for pure Zn \cite{muralidhara2011electrodeposition,tamurejo2023electrodeposited}. Additionally, the weak reflections at $2\theta$ $\approx 31.8^\circ$, $34.4^\circ$, and $36.2^\circ$ corresponded to the (100), (002), and (101) planes of wurtzite ZnO, indicating minor surface oxidation. However, the low intensity of these peaks indicated that the oxidation was limited to the surface due to exposure to atmospheric oxygen. The absence of any additional reflections confirmed the high phase purity of the sample, consistent with the reported literature \cite{nasui2025cold}.

%
%
%

In contrast, the incorporation of Al into Zn resulted in noticeable changes in the diffraction pattern, showing solid-solution formation and the emergence of intermetallic phases, as shown in Fig.~\ref{Fig05}. Although the primary Zn reflections remained dominant, their slight shifts toward higher $2\theta$ values presented lattice contraction due to the substitution of Zn atoms with smaller radius Al atoms. Additional reflections related to Al-Zn intermetallic compounds were observed in the $36^\circ$--$47^\circ$ and $65^\circ$--$69^\circ$ ranges. The peak broadening of the alloy sample compared to pure Zn indicated increased microstrain and reduced crystallite size. These changes may result from solid-solution strengthening mechanisms and the dispersion of intermetallics. For the Zn$_{0.9}$Al$_{0.1}$ alloy, the crystallite size and the microstrain were 20.21 nm and $4.1 \times 10^{-3}$, respectively. Similarly, the other Zn-Al alloy samples exhibited comparable XRD patterns with no clearly resolved additional diffraction peaks (see Figs.~S2--S4), indicating minimal structural variation across the compositions studied. Overall, the diffraction results supported a multiphase microstructure comprising Zn, Zn-Al intermetallic phases, and residual Al, confirming the formation of the Zn-Al alloys.

%
\begin{figure}
    \centering
    \includegraphics[width =0.71\linewidth]{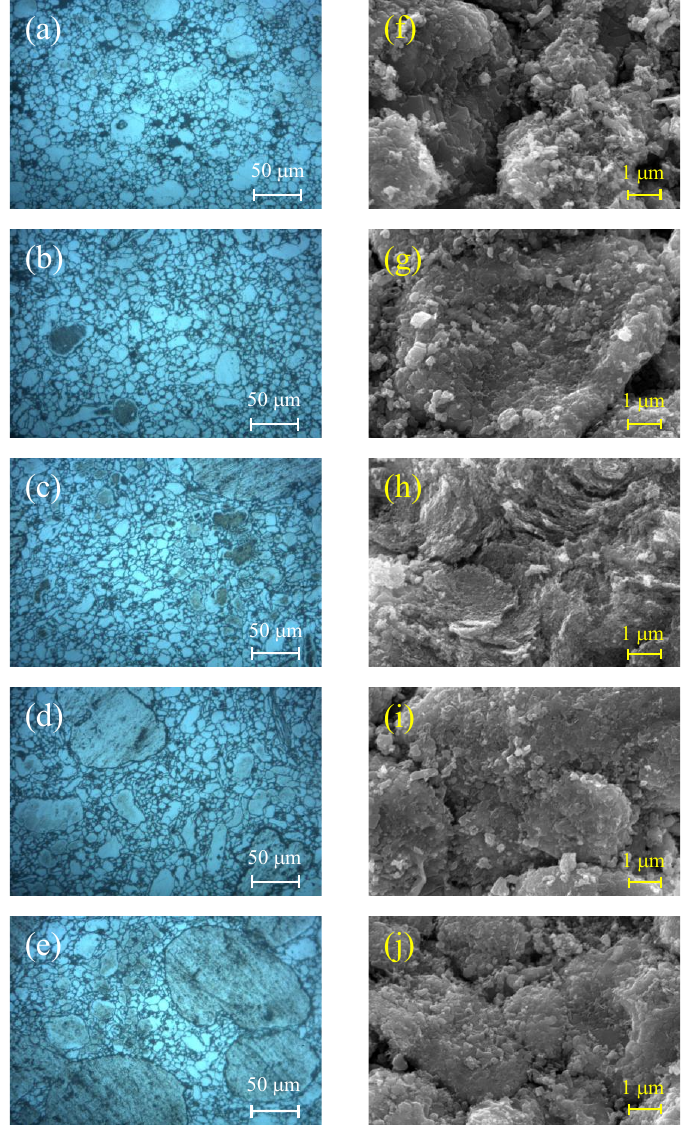}
    \caption{Optical Microscopy and scanning electron microscopy (SEM) images for (a),(f) Pure Zn, (b),(g) 5 wt.\% Al into Zn, (c),(h) 10 wt.\% Al into Zn, (d),(i) 15 wt.\% Al into Zn, (e),(j) 20 wt.\% Al into Zn, respectively.}
    \label{Fig06}
\end{figure}
%
%

The optical microscopy represented a distinct evolution of the microstructure in Zn-Al alloys in terms of increasing Al compositions, as shown in Figs.~\ref{Fig06}(a)--(e) and S2. In the pure Zn sample, the microstructure primarily consisted of a Zn matrix characterized by large grains and well-defined boundaries with no signs of phase segregation (see Fig.~\ref{Fig06}(a)), marking a homogeneous and non-interconnected Zn metallic phase. With the addition of 5 wt.\% Al, small $\alpha$-phase of Al-rich precipitates began to form preferentially along the grain boundaries, as illustrated in Fig.~\ref{Fig06}(b). This addition slightly refined the grains and introduced minor phase contrast. When the Al content reached 10 wt.\%, the Zn-Al alloy approached a near-eutectoid structure, showing a refined and interconnected mixture of Zn-rich $\eta$ and Al-rich $\alpha$ phases, as displayed in Fig.~\ref{Fig06}(c). Consequently, the grain boundaries became more continuous, and the phase distribution appeared more uniform.

Similarly, the Al content of 15 wt.\% resulted in a near-optimal eutectic composition, producing a well-distributed and interconnected multiphase network, as depicted in Fig.~\ref{Fig06}(d). In this case, the grains became finer, and the light and dark regions referred to Zn-rich and Al-rich domains, respectively, showing a more homogeneously dispersed pattern. When the Al content increases to 20 wt.\%, the microstructure exhibits a distinctly multiphasic character, featuring larger Al-rich regions embedded within the primary Zn matrix, as presented in Fig.~\ref{Fig06}(e). This behavior likely reflects the formation of coarser second-phase aggregates and larger Al-rich domains due to the deviation from the eutectic point and a solidification path. The coarsening observed at higher Al composition resulted in reduced uniformity, larger characteristic phase domains, and decreased structural refinement compared to the sample with 15 wt.\% Al. Therefore, the optical micrographs demonstrated that increasing Al up to 15 wt.\% led to a highly refined and strongly interconnected eutectic-eutectoid structure near the eutectic composition, which can enhance the material properties and electrochemical performance of Zn-Al alloys.

The SEM examination of Zn-Al alloys provided more detailed insights into the phase morphology, distribution, and interfaces at micro-to-submicron scales, as shown in Figs.~\ref{Fig06}(f)--(j) and S3. The pure Zn sample exhibited relatively large and irregularly shaped particles with noticeable inter-particle porosity. Incorporating Al into the Zn matrix resulted in finer and more uniform particles, demonstrating improved densification with increasing Al compositions. At 15 wt.\% Al, the SEM microstructure exhibited a highly uniform and fine-scale eutectoid-eutectic interconnected phases. Consequently, the boundaries between the phases were sharp, and the phase areas were small and had more distributed characteristics, offering superior mechanical strength and materials properties. However, when Al content reached 20 wt.\%, the SEM images showed a coarser and more inhomogeneous microstructure, including larger intermetallic grains and potential dendritic segregation. Furthermore, this coarsening led to decreased uniformity in phase distribution, which could decline mechanical homogeneity and result in poor material properties. In summary, both the SEM images and the optical microscopy findings demonstrated that Al facilitated grain refinement and densification up to 15 wt.\% Al compositions, while further incorporating Al resulted in microstructural coarsening and a loss of uniformity.

\subsection{Electrochemical Analysis}

\begin{figure}[hbtp]
    \centering
    \includegraphics[width =0.99\linewidth]{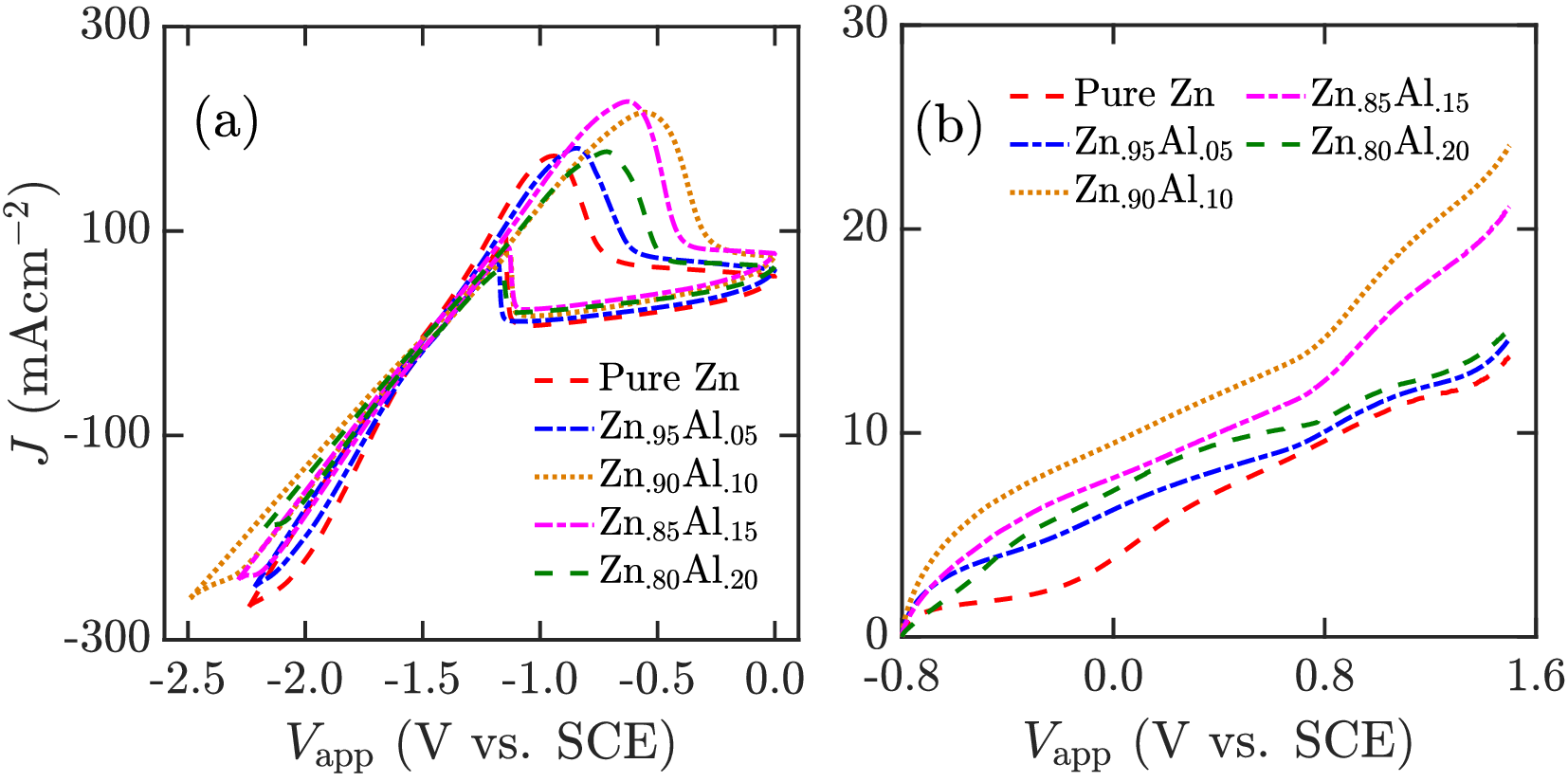}
    \caption{(a) Cyclic voltammetry (CV) and (b) polarization curve (PC) for Zn-Al alloy electrodes in 1 M KOH solution and at 25 $^\circ$C temperature.}
    \label{Fig07}
\end{figure}

Electrochemical measurements provide a comprehensive understanding of the mechanistic behaviors and performance of water-splitting processes, including potential-dependent kinetics, interfacial charge-transfer dynamics, $\eta_0$, and assessment of electrochemical stability. In our study, we performed CV on Zn-Al alloys to investigate their potential-dependent redox behavior, surface reconstructions, and activation of catalytic sites relevant to evolution reaction pathways. The pure Zn electrode exhibited distinct anodic and cathodic peaks, which correspond to the oxidation and reduction processes occurring at its surface, as shown in Fig.~\ref{Fig07}(a). However, the current densities were relatively low, implying that pure Zn limited electrochemical reactivity and charge-transfer kinetics. Furthermore, the peak for pure Zn occurred at a lower potential, indicating the formation of surface passivation layers, such as ZnO and Zn(OH)$_2$. These layers hindered electrocatalytic activity and efficient charge transfer. 

Conversely, the Zn-Al alloys displayed higher current responses and wider potential ranges compared to pure Zn, signifying improved performance in water-splitting processes, as illustrated in Fig.~\ref{Fig07}(a). This improvement resulted from the synergistic effects of Zn and Al atoms, which can modify surface morphology, enhance redox behavior, and increase electrical conductivity in the alloys. Notably, the increased anodic and cathodic currents observed for Al-rich alloys suggested improved kinetics for evolution reactions in water splitting. This enhancement stemmed from altered adsorption energetics of intermediate species and the potential formation of metastable intermetallic surface phases, which provided more reaction-active sites. 

%
%
%

The distinct anodic features observed between $-1.0$ and $-0.8$ V vs. SCE corresponded to the oxidation and subsequent re-reduction of mixed Zn-Al hydroxide-oxides formation. Among the various compositions analyzed, the Zn$_{0.9}$Al$_{0.1}$ and Zn$_{0.85}$Al$_{0.15}$ electrodes exhibited the highest anodic and cathodic current densities, respectively, enabling excellent electrochemical performance. However, as Al content increased to 20 wt.\%, $J$ began to decline. This degradation can be attributed to the partial passivation and reduced active sites resulting from the increased formation of Al$_2$O$_3$ films on the alloy surface. Additionally, our analysis revealed that Zn$_{0.9}$Al$_{0.1}$ and Zn$_{0.85}$Al$_{0.15}$ showed shifts in oxidation peaks toward more positive potentials and reduction peaks toward more negative potentials compared to pure Zn. These shifts suggest that these alloys can modify the thermodynamic characteristics of the redox processes, which can decrease the activation energy barriers for charge transfer at the electrode--electrolyte interface.

Figure \ref{Fig07}(b) illustrates the anodic polarization curves (PC) for pure Zn and Zn-Al alloys. These curves showed a steadily elevated $J$ with increasing Al contents across the applied potential ranges. The pure Zn electrode exhibited the lowest $J$ throughout the entire voltage range, indicating its limited catalytic activity. In contrast, the Zn-Al alloys, particularly Zn$_{0.9}$Al$_{0.1}$ and Zn$_{0.85}$Al$_{0.15}$, showed significantly higher $J$ compared to pure Zn. This advancement suggested that the presence of Al enhanced the electro-active surface area, increased the density of redox-accessible states, and improved electron transfer pathways in the modified reactive surface. The polarization curves for Zn$_{0.95}$Al$_{0.05}$ and Zn$_{0.8}$Al$_{0.2}$ nearly corresponded to the pure Zn, implying that moderate compositions of Al led to optimal surface modifications and exceptional materials properties. Overall, these outcomes emphasize the substantial impact of Zn-Al alloys on both the kinetics and structural evolution of reaction surfaces during the OER process.

\begin{figure}[hbtp]
    \centering
    \includegraphics[width =0.99\linewidth]{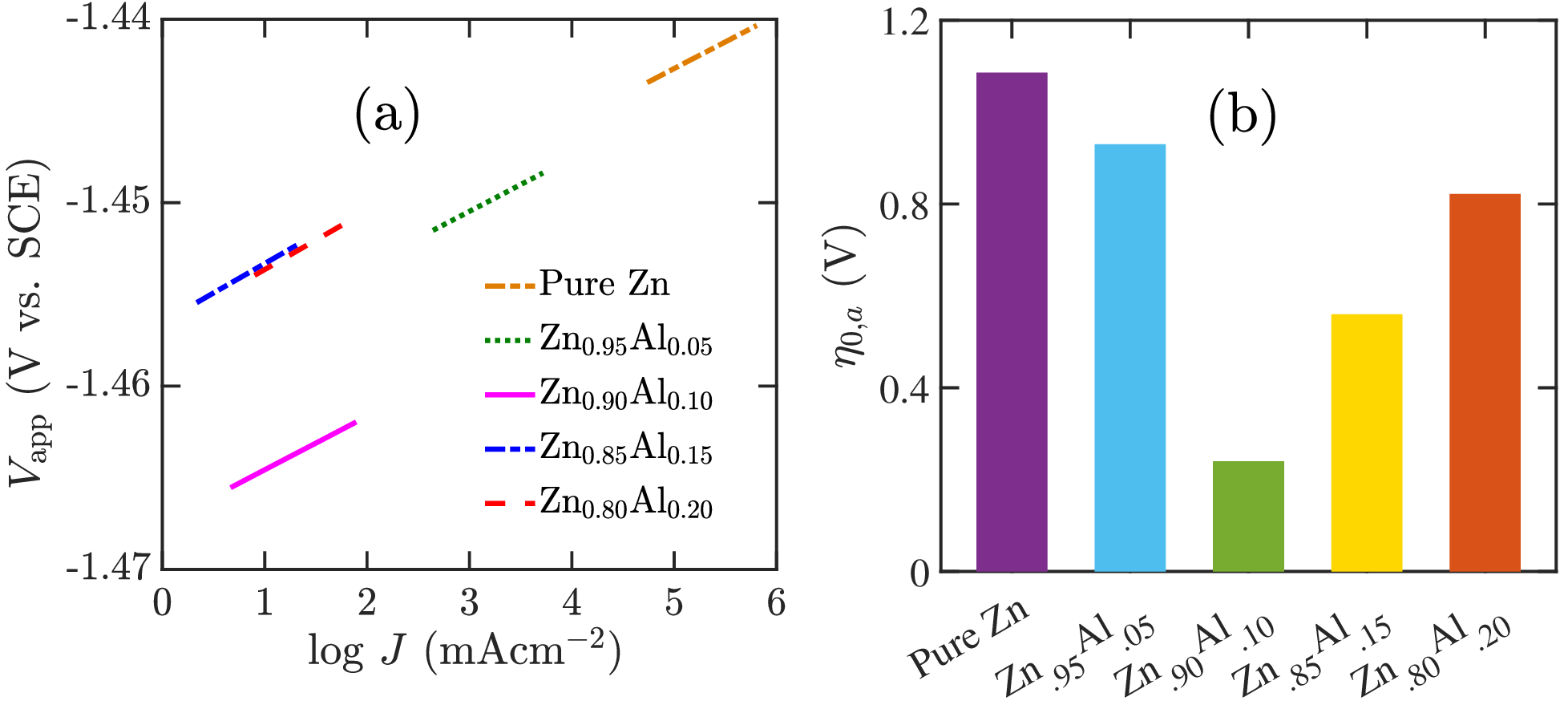}
    \caption{(a) Anodic Tafel slope and (b) anodic overpotential losses ($\eta_{0,a}$) at a current density ($J$) of 12 mAcm$^{-2}$ for Zn-Al alloy electrodes in 1 M KOH solution and at 25 $^\circ$C temperature.}
    \label{Fig08}
\end{figure}

The Tafel analysis and LPR techniques provide valuable insights into the charge-transfer kinetics and corrosion characteristics associated with the evolution reaction processes. A lower Tafel slope indicates faster electron-transfer kinetics and improved electrochemical performance. Figures \ref{Fig08}(a) and S5 illustrate the anodic Tafel slopes and LPR curves for the studied electrodes, respectively. The pure Zn possessed the highest Tafel slope at 87.9 mVdec$^{-1}$, indicating very slow anodic reaction kinetics. This behavior is typically related to the formation of passive layers of ZnO and Zn(OH)$_2$, which hinders electron transport, thereby limiting charge transport dynamics. The measured Tafel slopes for the Zn-Al alloys were 73.916, 53.491, 58.225, and 66.578 mVdec$^{-1}$ for Zn$_{0.95}$Al$_{0.05}$, Zn$_{0.9}$Al$_{0.1}$, Zn$_{0.85}$Al$_{0.15}$, and Zn$_{0.8}$Al$_{0.2}$, respectively. These results suggest that Zn$_{0.9}$Al$_{0.1}$ and Zn$_{0.85}$Al$_{0.15}$ exhibited an increase in kinetic efficiency of approximately 40\%, potentially due to electronic and structural modifications that can create more favorable charge-transfer pathways. Conversely, the higher Al content of 20 wt.\% in Zn induced excessive formation of Al-rich oxide layers, which reduced the number of available active sites, consequently diminishing anodic kinetic efficiency.

%
%
%

The $E_{\rm corr}$, charge-transfer coefficient ($\alpha_{n,a}$), and $J_{0,a}$ were derived from the Tafel slopes and LPR curves, as presented in Table \ref{Table2}. The changes in $E_{\rm corr}$ and $R_p$ were minimal compared to pure Zn, indicating that the corrosion behavior of the Zn-Al alloys was not significantly affected by Al incorporation and remained stable, comparable to that of pure Zn in a 1 M KOH electrolyte environment. The $\alpha_{n,a}$ reflects the symmetry of the energy barrier for electrochemical reactions, typically ranging from $0$ to $1$. A value of $\alpha_{n,a} = 0.5$ represents a perfectly symmetrical energy barrier. A higher $\alpha_{n,a}$, particularly values close to 0.5, corresponds to a more efficient reaction across the applied potential. Conversely, a lower $\alpha_{n,a}$ signifies an asymmetric activation barrier and less favorable for evolution reactions. Among all compositions, Zn$_{0.9}$Al$_{0.1}$ and Zn$_{0.85}$Al$_{0.15}$ exhibited the $\alpha_{n,a}$ values of $ 0.4803$ and $0.4412$, respectively, close to $0.5$, showing the most rapid potential-driven charge transfer characteristics and resulting in superior electrocatalytic performance. 

%
%

\begin{table}[hbt]
\centering
\caption{Anodic Tafel slope ($b_a$), corrosion potential ($E_{\rm corr}$), charge transfer coefficient ($\alpha_{n,a}$), anodic exchange current density ($J_{0,a}$), polarization resistance ($R_p$) and charge-transfer resistance ($R_{\rm CT}$) for Zn-Al alloy electrodes in 1 M KOH solution and at 25 $^\circ$C temperature.}
\resizebox{0.85\textwidth}{!}{%
\begin{tabular}{c c c c c c c}
\Xhline{3\arrayrulewidth}
    Electrode & $b_{\rm a}$ & $E_{\rm corr}$ & $\alpha_{\rm n}$ & $J_{0,a}$ & $R_{\rm p}$ & $R_{\rm CT}$ \\ 
    Name & (mVdec$^{-1}$) & (V$_{\rm SCE}$) & -- & (Acm$^{-2}$) & ($\Omega {\rm cm}^2$) & ($\Omega {\rm cm}^2$) \\
    \Xhline{2\arrayrulewidth}
    Pure Zn                & $87.911$  & $-1.4799$ & $0.2922$ & $3.4992 \times 10^{-5}$ & $2.9157$ & $9.52$  \\
    Zn$_{0.95}$Al$_{0.05}$ & $73.916$  & $-1.4804$ & $0.3475$ & $4.5868 \times 10^{-5}$ & $2.9192$ & $3.57$  \\
    Zn$_{0.90}$Al$_{0.10}$ & $53.491$  & $-1.4895$ & $0.4803$ & $13.7592 \times 10^{-5}$ & $2.9221$ & $1.85$  \\
    Zn$_{0.85}$Al$_{0.15}$ & $58.225$  & $-1.4784$ & $0.4412$ & $7.6328 \times 10^{-5}$ & $3.1991$ & $1.76$  \\
    Zn$_{0.80}$Al$_{0.20}$ & $66.578$  & $-1.4810$ & $0.3859$ & $6.8087 \times 10^{-5}$ & $3.2183$ & $2.13$  \\
    \Xhline{3\arrayrulewidth}
\end{tabular}}
\label{Table2}
\end{table}
%
%

The calculated $J_{0,a}$ values for Zn$_{0.9}$Al$_{0.1}$ and Zn$_{0.85}$Al$_{0.15}$ were $13.7592 \times 10^{-5}$  and $7.6328 \times 10^{-5}$ Acm$^{-2}$, respectively, both significantly higher than that of pure Zn, which had $J_{0,a} = 3.4992 \times 10^{-5}$ Acm$^{-2}$. These enhancements indicate faster reaction kinetics with reduced overpotential losses. Similarly, Zn$_{0.95}$Al$_{0.05}$ and Zn$_{0.8}$Al$_{0.2}$ exhibited higher $J_{0,a}$ values compared to pure Zn, although still lower than those of Zn$_{0.9}$Al$_{0.1}$ and Zn$_{0.85}$Al$_{0.15}$, demonstrating a moderate improvement in charge-transfer dynamics at the electrode--electrolyte interfaces.

Figure \ref{Fig08}(b) illustrates the $\eta_{0,a}$ of pure Zn and Zn-Al alloys with varying Al compositions at $J = 12$ mAcm$^{-2}$. For pure Zn, the value of $\eta_{0,a}$ was $1.0865$ V, indicating its poor electrocatalytic performance due to several inherent characteristics, including higher activation energy, a limited number of active sites, and lower electrical conductivity. In contrast, the addition of Al led to a decrease in $\eta_{0,a}$. The Zn$_{0.95}$Al$_{0.05}$ showed a reduced $\eta_{0,a}$ of $0.9309$ V, showing improved charge transfer compared to pure Zn. Notably, the lowest $\eta_{0,a}$ value of $0.2408$ V was observed for Zn$_{0.9}$Al$_{0.10}$, signifying that Al content of 10 wt.\% achieved optimal microstructural and electrochemical conditions in water splitting. On the other hand, Zn$_{0.85}$Al$_{0.15}$ and Zn$_{0.8}$Al$_{0.2}$ exhibited the $\eta_{0,a}$ values of $0.5603$ and $0.8221$ V, respectively. This increase in $\eta_{0,a}$ can result from the introduction of detrimental secondary phases and structural disruptions, which negatively impact electrochemical activity. This comprehensive analysis highlights the optimized composition of Al into Zn-Al alloys to minimize $\eta_0$ and significantly enhance energy efficiency in electrochemical applications.

\begin{figure}[H]
    \centering
    \includegraphics[width =0.97\linewidth]{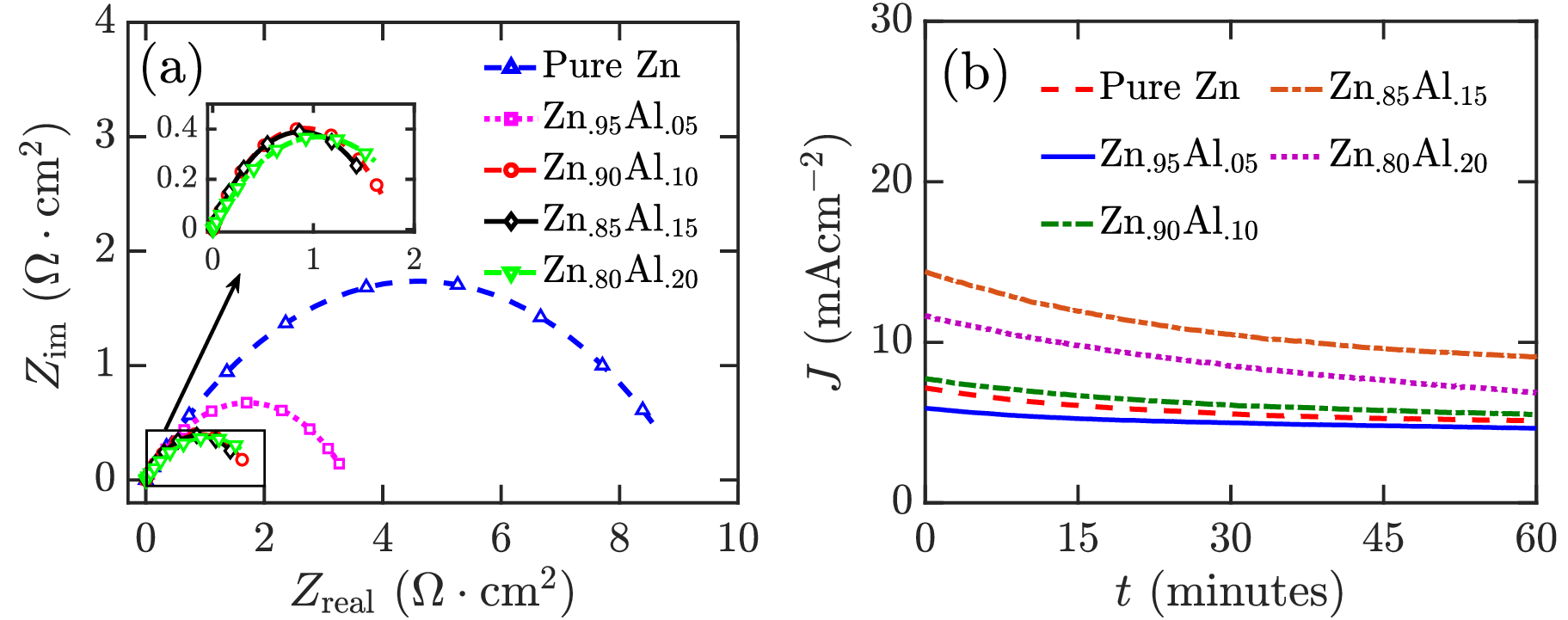}
    \caption{(a) Electrochemical impedance spectroscopy (EIS) and (b) Chronoamperometry analyses for pure Zn and Zn-Al alloy-based electrodes in 1 M KOH solution and at 25 $^\circ$C temperature. The applied voltage ($V_{\rm app}$) for chronoamperometry was $+0.8$ V vs. saturated calomel electrode (SCE).}
    \label{Fig09}
\end{figure}

We conducted EIS analysis to assess the $R_{\rm CT}$ of pure Zn and Zn-Al alloys, as shown in Fig.~\ref{Fig09}(a). The $R_{\rm CT}$ value is derived from the diameter of the Nyquist semicircle and is inversely related to the rate of charge transfer. A high $R_{\rm CT}$ indicates slow electron transfer processes and low catalytic activity, while a low $R_{\rm CT}$ offers faster reaction kinetics and enhanced catalytic efficiency. For pure Zn, the measured $R_{\rm CT}$ was at 9.52 $\Omega$cm$^{2}$, which aligns with previously reported values in the literature \cite{el2012corrosion}. In comparison, the Zn-Al alloys exhibited significantly lower $R_{\rm CT}$ values than pure Zn, indicating increased surface active sites, improved electronic structures, and rapid charge-transfer kinetics. The measured $R_{\rm CT}$ values for the Zn$_{0.95}$Al$_{0.05}$, Zn$_{0.9}$Al$_{0.1}$, Zn$_{0.85}$Al$_{0.15}$, and Zn$_{0.8}$Al$_{0.2}$ alloys were at 3.57, 1.85, 1.76, and 2.13 $\Omega$cm$^{2}$, respectively, all of which were significantly lower than that of pure Zn. Noticeably, the $R_{\rm CT}$ values for both Zn$_{0.9}$Al$_{0.1}$ and Zn$_{0.85}$Al$_{0.15}$ were close and the lowest among the studied alloys, facilitating rapid charge transport across the electrode--electrolyte interfaces. Overall, these results clearly demonstrate that the incorporation of Al in Zn significantly accelerated interfacial charge-transfer processes, with Zn$_{0.9}$Al$_{0.1}$ and Zn$_{0.85}$Al$_{0.15}$ exhibiting the most efficient electron transport.

%
%
%

To assess the stability, we conducted chronoamperometry analysis on both pure Zn and Zn-Al alloys at $V_{\rm app} = +0.8$ V vs.~SCE for a duration of $60$ minutes, as shown in Fig.~\ref{Fig09}(b). The chronoamperometry technique provides insights into both the chemical stability and the active catalytic performance under a constant $V_{\rm app}$, where a slower decay and flatter slopes in $J$ indicate greater stability and sustained reactivity over time. The Zn$_{0.85}$Al$_{0.15}$ exhibited the highest initial $J$ and maintained a significantly elevated response throughout the measurement period. This behavior may result from favorable interfacial charge-transfer characteristics, suggesting a high density of active sites and efficient electron-transport pathways. Although the reaction kinetic rate was slower for the Zn$_{0.8}$Al$_{0.2}$, it still displayed a relatively high $J$, which was smaller than that of Zn$_{0.85}$Al$_{0.15}$ but greater than that of Zn$_{0.9}$Al$_{0.1}$. However, this characteristic does not necessarily signify improved catalytic performance. Such a high Al content can lead to the formation of heterogeneous phases and Al-rich oxide layers, driven by various kinetic and corrosion-related processes rather than purely intrinsic catalytic activity in water splitting, resulting in an elevated $J$. 

In contrast, the Zn$_{0.9}$Al$_{0.1}$ showed a smaller but notably stable $J$ of approximately $7$--$8$ mAcm$^{-2}$, corresponding to its exceptionally rapid charge-transfer kinetics. This relatively modest $J$ suggests highly controlled surface reactions with minimized parasitic processes, reflecting the alloy's optimal balance between electronic conductivity and structural uniformity. For the Zn$_{0.95}$Al$_{0.05}$ and pure Zn samples, the current densities remained lower, indicating limited electrochemical activation associated with slower interfacial kinetics. Overall, this comprehensive investigation suggests that alloys with 10 wt.\% and 15 wt.\% Al provided the most kinetically favorable compositions, while the elevated response of the 20 wt.\% aluminum alloy likely reflected secondary effects rather than enhanced intrinsic electrochemical activity.

\begin{table}[hbt]
\centering
\caption{Anodic Tafel slope ($b_a$), anodic overpotential loss ($\eta_{0,a}$), and charge-transfer resistance ($R_{\rm CT}$) for different electrodes.}
\resizebox{0.76\textwidth}{!}{%
\begin{tabular}{c c c c c}
\Xhline{3\arrayrulewidth}
    Electrode & $b_{\rm a}$ & $\eta_{0,a}$ & $R_{\rm CT}$ & Ref. \\ 
    Name & (mVdec$^{-1}$) & (V) & ($\Omega {\rm cm}^2$) & --  \\
    \Xhline{2\arrayrulewidth}
    Ni-Fe                            & $105$        & $0.350^f$           & $2.39$   & \cite{elsharkawy2025effect}  \\
    Co$_3$O$_4$                      & $70$--$79$   & $0.355$--$0.384^f$  & $5.365$  & \cite{kim2023electrocatalytic}   \\
    Mn$_2$CoO$_4$                    & $78.9$       & $0.399^f$           & $1.52$   & \cite{lankauf2020mnxco3}  \\
    FeNiHOF                          & $34.8$       & $0.273^f$           & $1.23$   & \cite{chen2024boosting}  \\
    rGO-NiFe(83:17)                  & $84.4$       & $0.420^f$           & $3.10$   & \cite{garcia2025nife}  \\
    Fe$_3$O$_4$@Co$_9$S$_8$/rGO      & $144$        & $0.340^f$           & --       & \cite{yang2016fe3o4}   \\
    NP Co$_3$O$_4$/Fe@C$_2$N	     & $103$        & $0.432^f$           & $2.5$    & \cite{kim2019synergistic}  \\          
    Zn$_{0.90}$Al$_{0.10}$           & $53.491$     & $0.240^g$           & $1.85$   & This work  \\
    Zn$_{0.85}$Al$_{0.15}$           & $58.225$     & $0.560^g$           & $1.76$   & This work  \\    
    \Xhline{3\arrayrulewidth}
    \multicolumn{5}l {{$^f$ and $^g$ represent the measured $\eta_{0,a}$ at $J$ of 10 and 12 mAcm$^{-2}$, respectively.}} 
\end{tabular}}
\label{Table3}
\end{table}
%
%

To assess the efficiency and effectiveness of OER catalysts, it is essential to compare their kinetic parameters and catalytic activities. Table \ref{Table3} represents a comparison of Zn-Al alloys against various catalysts for the OER in 1 M KOH solution. The Zn-Al alloy electrodes demonstrated significantly improved OER activity compared to a wide range of reported catalysts, as highlighted by key anodic kinetic parameters. Notably, the Zn$_{0.9}$Al$_{0.1}$ alloy exhibited a $b_a$ value of 53.49 mVdec$^{-1}$, considerably smaller than those of Ni-Fe, Co$_3$O$_4$, Mn$_2$CoO$_4$, and rGO-NiFe. This smaller $b_a$ value indicates rapid charge transfer at the electrode--electrolyte interface, which is crucial for determining the reaction kinetic rate. Furthermore, the Zn$_{0.9}$Al$_{0.1}$ alloy showed $\eta_{\rm 0,a}$ of $0.240$ V at $J$ of 12 mAcm$^{-2}$, surpassing advanced complex catalysts, such as FeNiHOF, rGO-NiFe(83:17), and Fe$_3$O$_4$@Co$_9$S$_8$/rGO. The Zn$_{0.85}$Al$_{0.15}$ alloy also displayed favorable kinetics, with a low $b_a$ of 58.23 mVdec$^{-1}$ and the lowest $R_{\rm CT}$ of 1.76 $\Omega$cm$^2$, indicating highly efficient interfacial charge transport. Although its overpotential was slightly higher, its kinetic metrics remained superior to most oxide-based electrodes. In summary, Zn-Al alloy electrodes outperform or are at least comparable to the reported OER catalysts, highlighting that Zn-Al alloys have strong potential for efficient and economically viable water-splitting applications.

Finally, both theoretical and experimental studies emphasize the importance of optimizing the Al composition in Zn-Al alloys. The Zn-Al phase diagram indicates that variations in Al content affect the phase composition and microstructure, ultimately influencing the electronic structures and surface chemistry necessary for catalysis \cite{durmus2019influence}. The optimal Al concentrations of 10 wt.\% and 15 wt.\% into Zn exhibited excellent reaction kinetics and charge-transfer dynamics, which significantly reduced $\eta_{0,a}$ and enhanced catalytic efficiency. In these compositions, the alloys remain in a solid-solution or near-hypoeutectic state, suggesting that microstructural refinement and lattice distortions improve conductivity and increase the number of active sites. Although the content of 5 wt.\% Al into Zn showed superior electronic properties, but its catalytic activity presented only marginal improvements due to insufficient active sites and limited microstructural modifications. This small amount of Al typically leads to minimal changes in phase morphology and defect density, resulting in only slight improvements in electrochemical performance despite modest improvements in electronic characteristics.

On the other hand, at 20 wt.\% Al resulted in the segregation of metastable secondary phases within the Zn-Al alloy. This higher Al content, which exceeds the primary solid-solution region of the Zn-Al phase diagram, is prone to the formation of Al-rich and eutectic constituents that can disrupt overall structural homogeneity. While the Zn$_{0.8}$Al$_{0.2}$ alloy demonstrated moderately improved catalytic activity compared to pure Zn, it also produced Al-rich oxides and promoted corrosion, limiting its application in sustainable water-splitting systems. As a consequence, the formation of Al-rich regions and oxides at higher Al concentrations negatively impacts the corrosion behavior of Zn-Al alloy by creating electrochemical heterogeneity and altering surface reactions, which can be detrimental under alkaline conditions. Conversely, Zn$_{0.9}$Al$_{0.1}$ and Zn$_{0.85}$Al$_{0.15}$ emerge as the most promising electrodes among the Zn-Al alloys. These intermediate compositions maintain a balance between phase stability and enhanced electronic and microstructural properties while avoiding excessive phase segregation. Notably, the reaction kinetics and catalytic performance of Zn$_{0.9}$Al$_{0.1}$ were comparable to those of multifaceted and complex functional electrocatalysts, while still allowing for low-cost and simple alloy manufacturing processes, making them more effective and efficient for water splitting. These findings demonstrate that a carefully optimized Zn-Al alloy can achieve high catalytic performance while facilitating scalable and economical fabrication, underscoring its practical advantages over more complex electrocatalysts in alkaline water electrolysis.

%
%
\section{Conclusion}

In summary, we developed binary Zn-Al alloys with Al compositions up to 20 wt.\% using both theoretical and experimental methods to identify the optimal composition. The incorporation of 5 wt.\% Al in Zn resulted in excellent thermodynamic and electronic properties, while the limited availability of reactive sites hindered the kinetic rates and catalytic efficiency. In contrast, the Al content of 20 wt.\% into Zn exhibited metastable structural features and Al-rich surface oxidation, which obstructed charge-transport pathways and ultimately led to poor electrochemical performance. Notably, the Zn-Al alloys with 10 wt.\% and 15 wt.\% Al demonstrated outstanding structure characteristics, electronic properties, and catalytic activity. The $J_{0,a}$ for Zn$_{0.9}$Al$_{0.1}$ and Zn$_{0.85}$Al$_{0.15}$ were measured at $13.7592 \times 10^{-5}$ and $7.6328 \times 10^{-5}$ Acm$^{-2}$, respectively, both significantly higher than that of pure Zn, which was $3.4992 \times 10^{-5}$ Acm$^{-2}$. Additionally, the $R_{\rm CT}$ for these alloys was nearly four times lower than that of pure Zn, indicating rapid reaction rates and efficient charge transport at the electrode--electrolyte interface. These findings highlight the potential of these alloys as low-cost, high-performance electrodes for applications not only in water electrolysis but also in batteries and electronic devices.

%
%

\section*{Supplementary Material}
The supplementary material presents the designed Zn-Al alloy crystal structures, X-Ray diffraction (XRD) pattern, and alternative resolution optical microscopy and scanning electron microscopy (SEM) images, as well as potentiodynamic linear polarization curves (LPR) for both pure and Zn-Al alloy electrodes.. 

\section*{Data Availability}
All data of the paper are presented in the main text and the supplementary material.

%
\section*{Author Declaration} 
The authors have no conflicts to disclose.

%
\section*{Acknowledgments} 
The authors gratefully acknowledge financial support from the Research and Innovation Centre for Science and Engineering (RISE) at Bangladesh University of Engineering and Technology (BUET). The authors also acknowledge experimental and computational assistance received from the Photonics Laboratory in the Department of Electrical and Electronic Engineering (EEE) and the Foundry Laboratory in the Department of Materials and Metallurgical Engineering at BUET.

%

%
%
\small
\bibliographystyle{ieeetr}
\bibliography{references}

@article{mamun2025advancing,
  title={Advancing Transition Metal Oxide Photoelectrodes for Efficient Solar-Driven Hydrogen Generation: Strategies and Insights},
  author={Mamun, Abdul Ahad and Chowdhury, Abir Hasan and Billah, Asif and Karim, Jawadul and Hussain, Auronno Ovid and Rahman, Faysal and Talukder, Muhammad Anisuzzaman},
  journal={Advanced Energy Materials},
  volume={15},
  number={32},
  pages={2501766},
  year={2025},
  publisher={Wiley Online Library}
}

@article{pitsch2024transition,
  title={The transition to sustainable combustion: Hydrogen-and carbon-based future fuels and methods for dealing with their challenges},
  author={Pitsch, Heinz},
  journal={Proceedings of the Combustion Institute},
  volume={40},
  number={1-4},
  pages={105638},
  year={2024},
  publisher={Elsevier}
}

@article{mamun2024techno,
  title={Techno-economic analysis of the direct solar conversion of carbon dioxide into renewable fuels},
  author={Mamun, Abdul Ahad and Talukder, Muhammad Anisuzzaman},
  journal={Energy Conversion and Management},
  volume={321},
  pages={119038},
  year={2024},
  publisher={Elsevier}
}

@article{fearnside2025cop,
  title={\textsc{COP}30: Brazilian policies must change},
  author={Fearnside, Philip M and Filho, Walter Leal},
  journal={Science},
  volume={387},
  number={6740},
  pages={1237--1237},
  year={2025},
  publisher={American Association for the Advancement of Science}
}

@article{mamun2024enhancing,
  title={Enhancing hydrogen evolution reaction using iridium atomic monolayer on conventional electrodes: A first-principles study},
  author={Mamun, Abdul Ahad and Billah, Asif and Talukder, Muhammad Anisuzzaman},
  journal={International Journal of Hydrogen Energy},
  volume={59},
  pages={982--990},
  year={2024},
  publisher={Elsevier}
}

@article{segovia2025green,
  title={Green hydrogen production for sustainable development: a critical examination of barriers and strategic opportunities},
  author={Segovia-Hern{\'a}ndez, Juan Gabriel and Hern{\'a}ndez, Salvador and Coss{\'\i}o-Vargas, Enrique and Juarez-Garc{\'\i}a, Maricruz and S{\'a}nchez-Ram{\'\i}rez, Eduardo},
  journal={RSC Sustainability},
  volume={3},
  number={1},
  pages={134--157},
  year={2025},
  publisher={Royal Society of Chemistry}
}

@article{qian2024recent,
  title={Recent advancements in electrochemical hydrogen production via hybrid water splitting},
  author={Qian, Qizhu and Zhu, Yin and Ahmad, Nazir and Feng, Yafei and Zhang, Huaikun and Cheng, Mingyu and Liu, Huanhuan and Xiao, Chong and Zhang, Genqiang and Xie, Yi},
  journal={Advanced Materials},
  volume={36},
  number={4},
  pages={2306108},
  year={2024},
  publisher={Wiley Online Library}
}

@article{dutta2018designing,
  title={Designing electrochemically reversible \textsc{H}$_2$ oxidation and production catalysts},
  author={Dutta, Arnab and Appel, Aaron M and Shaw, Wendy J},
  journal={Nature Reviews Chemistry},
  volume={2},
  number={9},
  pages={244--252},
  year={2018},
  publisher={Nature Publishing Group UK London}
}

@article{rahman2025enhanced,
  title={Enhanced Hydrogen Evolution Using $\beta$-\textsc{M}n\textsc{O}$_2$ Monolayer on \textsc{N}i Electrode with Engineered Oxygen Vacancies},
  author={Rahman, Faysal and Mamun, Abdul Ahad and Hussain, Auronno Ovid and Talukder, Muhammad Anisuzzaman},
  journal={The Journal of Physical Chemistry C},
  volume={129},
  number={38},
  pages={17049--17060},
  year={2025},
  publisher={ACS Publications}
}

@article{ruan2025technologies,
  title={Technologies and prospects for decoupled and membraneless water electrolysis},
  author={Ruan, Guilin and Todman, Fiona and Yogev, Gilad and Arad, Rotem and Smolinka, Tom and Jensen, Jens Oluf and Symes, Mark D and Rothschild, Avner},
  journal={Nature Reviews Clean Technology},
  pages={1--16},
  year={2025},
  publisher={Nature Publishing Group UK London}
}

@article{wan2023key,
  title={Key components and design strategy of the membrane electrode assembly for alkaline water electrolysis},
  author={Wan, Lei and Xu, Ziang and Xu, Qin and Pang, Maobing and Lin, Dongcheng and Liu, Jing and Wang, Baoguo},
  journal={Energy \& Environmental Science},
  volume={16},
  number={4},
  pages={1384--1430},
  year={2023},
  publisher={Royal Society of Chemistry}
}

@article{mamun2023effects,
  title={Effects of activation overpotential in photoelectrochemical cells considering electrical and optical configurations},
  author={Mamun, Abdul Ahad and Billah, Asif and Talukder, Muhammad Anisuzzaman},
  journal={Heliyon},
  volume={9},
  number={6},
  year={2023},
  publisher={Elsevier}
}

@article{exner2019beyond,
  title={Beyond the rate-determining step in the oxygen evolution reaction over a single-crystalline \textsc{I}r\textsc{O}$_2$ (110) model electrode: kinetic scaling relations},
  author={Exner, Kai S and Over, Herbert},
  journal={ACS Catalysis},
  volume={9},
  number={8},
  pages={6755--6765},
  year={2019},
  publisher={ACS Publications}
}

@article{yao2021strategy,
  title={A strategy for preparing high-efficiency and economical catalytic electrodes toward overall water splitting},
  author={Yao, Dongxue and Gu, Lingling and Zuo, Bin and Weng, Shuo and Deng, Shengwei and Hao, Weiju},
  journal={Nanoscale},
  volume={13},
  number={24},
  pages={10624--10648},
  year={2021},
  publisher={Royal Society of Chemistry}
}

@article{hussain2025physics,
  title={Physics-informed electrochemical model of cathodic corrosion in alkaline media},
  author={Hussain, Auronno Ovid and Mamun, Abdul Ahad and Rahman, Faysal and Talukder, Muhammad Anisuzzaman},
  journal={Electrochimica Acta},
  volume={546},
  pages={147800},
  year={2025},
  publisher={Elsevier}
}

@article{shi2019robust,
  title={Robust noble metal-based electrocatalysts for oxygen evolution reaction},
  author={Shi, Qiurong and Zhu, Chengzhou and Du, Dan and Lin, Yuehe},
  journal={Chemical Society Reviews},
  volume={48},
  number={12},
  pages={3181--3192},
  year={2019},
  publisher={Royal Society of Chemistry}
}

@article{qin2024ru,
  title={\textsc{R}u/\textsc{I}r-based electrocatalysts for oxygen evolution reaction in acidic conditions: From mechanisms, optimizations to challenges},
  author={Qin, Rong and Chen, Guanzhen and Feng, Xueting and Weng, Jiena and Han, Yunhu},
  journal={Advanced Science},
  volume={11},
  number={21},
  pages={2309364},
  year={2024},
  publisher={Wiley Online Library}
}

@article{yan2016review,
  title={A review on noble-metal-free bifunctional heterogeneous catalysts for overall electrochemical water splitting},
  author={Yan, Ya and Xia, Bao Yu and Zhao, Bin and Wang, Xin},
  journal={Journal of Materials Chemistry A},
  volume={4},
  number={45},
  pages={17587--17603},
  year={2016},
  publisher={Royal Society of Chemistry}
}

@article{datye2021opportunities,
  title={Opportunities and challenges in the development of advanced materials for emission control catalysts},
  author={Datye, Abhaya K and Votsmeier, Martin},
  journal={Nature Materials},
  volume={20},
  number={8},
  pages={1049--1059},
  year={2021},
  publisher={Nature Publishing Group UK London}
}

@article{asefa2021nanostructured,
  title={Nanostructured carbon electrocatalysts for energy conversions},
  author={Asefa, Tewodros and Tang, Chaoyun and Ram{\'\i}rez-Hern{\'a}ndez, Maricely},
  journal={Small},
  volume={17},
  number={48},
  pages={2007136},
  year={2021},
  publisher={Wiley Online Library}
}

@article{roy2024engineered,
  title={Engineered two-dimensional transition metal dichalcogenides for energy conversion and storage},
  author={Roy, Soumyabrata and Joseph, Antony and Zhang, Xiang and Bhattacharyya, Sohini and Puthirath, Anand B and Biswas, Abhijit and Tiwary, Chandra Sekhar and Vajtai, Robert and Ajayan, Pulickel M},
  journal={Chemical Reviews},
  volume={124},
  number={16},
  pages={9376--9456},
  year={2024},
  publisher={ACS Publications}
}

@article{das2022transition,
  title={Transition metal non-oxides as electrocatalysts: advantages and challenges},
  author={Das, Chandni and Sinha, Nibedita and Roy, Poulomi},
  journal={Small},
  volume={18},
  number={28},
  pages={2202033},
  year={2022},
  publisher={Wiley Online Library}
}

@article{noor2021recent,
  title={Recent advances in electrocatalysis of oxygen evolution reaction using noble-metal, transition-metal, and carbon-based materials},
  author={Noor, Tayyaba and Yaqoob, Lubna and Iqbal, Naseem},
  journal={ChemElectroChem},
  volume={8},
  number={3},
  pages={447--483},
  year={2021},
  publisher={Wiley Online Library}
}

@article{edao2024nickel,
  title={Nickel--Iron Layered Double Hydroxides/Nickel Sulfide Heterostructured Electrocatalysts on Surface-Modified \textsc{T}i Foam for the Oxygen Evolution Reaction},
  author={Edao, Habib Gemechu and Chang, Chia-Yu and Dilebo, Woldesenbet Bafe and Angerasa, Fikiru Temesgen and Moges, Endalkachew Asefa and Nikodimos, Yosef and Guta, Chemeda Barasa and Lakshmanan, Keseven and Chen, Jeng-Lung and Tsai, Meng-Che and others},
  journal={ACS Applied Materials \& Interfaces},
  volume={16},
  number={38},
  pages={50602--50613},
  year={2024},
  publisher={ACS Publications}
}

@article{zhu2023improving,
  title={Improving the oxygen evolution activity of layered double-hydroxide via erbium-induced electronic engineering},
  author={Zhu, Yu and Wang, Xuan and Zhu, Xiaoheng and Wu, Zixin and Zhao, Dongsheng and Wang, Fei and Sun, Dongmei and Tang, Yawen and Li, Hao and Fu, Gengtao},
  journal={Small},
  volume={19},
  number={5},
  pages={2206531},
  year={2023},
  publisher={Wiley Online Library}
}

@article{singh2025zinc,
  title={Zinc-based materials for electrocatalytic reduction reactions: progress and prospects},
  author={Singh, Baghendra and Draksharapu, Apparao},
  journal={Materials Chemistry Frontiers},
  volume  ={9},
  issue  ={15},
  pages  ={2287-2321},
  year={2025},
  publisher={Royal Society of Chemistry}
}

@article{hao2020deeply,
  title={Deeply understanding the \textsc{Z}n anode behaviour and corresponding improvement strategies in different aqueous Zn-based batteries},
  author={Hao, Junnan and Li, Xiaolong and Zeng, Xiaohui and Li, Dan and Mao, Jianfeng and Guo, Zaiping},
  journal={Energy \& Environmental Science},
  volume={13},
  number={11},
  pages={3917--3949},
  year={2020},
  publisher={Royal Society of Chemistry}
}

@article{singh2025recent,
  title={Recent progress and advancement on zinc-based materials for water splitting: Structure-property-performance correlation},
  author={Singh, Baghendra and Draksharapu, Apparao},
  journal={Coordination Chemistry Reviews},
  volume={535},
  pages={216647},
  year={2025},
  publisher={Elsevier}
}

@article{chen2024work,
  title={Work function-guided electrocatalyst design},
  author={Chen, Zhijie and Ma, Tianyi and Wei, Wei and Wong, Wai-Yeung and Zhao, Chuan and Ni, Bing-Jie},
  journal={Advanced Materials},
  volume={36},
  number={29},
  pages={2401568},
  year={2024},
  publisher={Wiley Online Library}
}

@article{neyerlin2007study,
  title={Study of the exchange current density for the hydrogen oxidation and evolution reactions},
  author={Neyerlin, KC and Gu, Wenbin and Jorne, Jacob and Gasteiger, Hubert A},
  journal={Journal of The Electrochemical Society},
  volume={154},
  number={7},
  pages={B631},
  year={2007},
  publisher={IOP Publishing}
}

@article{nuss2010structural,
  title={The Structural Anomaly of Zinc: Evolution of Lattice Constants and Parameters of Thermal Motion in the Temperature Range of 40 to 500 \textsc{K}},
  author={Nuss, J{\"u}rgen and Wedig, Ulrich and Kirfel, Armin and Jansen, Martin},
  journal={Z. Anorg. Allg. Chem},
  volume={636},
  pages={309--313},
  year={2010}
}

@article{wedig2007structural,
  title={Structural and electronic properties of \textsc{M}g, \textsc{Z}n, and \textsc{C}d from Hartree-Fock and density functional calculations including hybrid functionals},
  author={Wedig, Ulrich and Jansen, Martin and Paulus, Beate and Rosciszewski, Krzysztof and Sony, Priya},
  journal={Physical Review B—Condensed Matter and Materials Physics},
  volume={75},
  number={20},
  pages={205123},
  year={2007},
  publisher={APS}
}

@article{tang2009surface,
  title={Surface structure and solidification morphology of aluminum nanoclusters},
  author={Tang, FL and Che, XX and Lu, WJ and Chen, GB and Xie, Y and Yu, WY},
  journal={Physica B: Condensed Matter},
  volume={404},
  number={16},
  pages={2489--2494},
  year={2009},
  publisher={Elsevier}
}

@article{olguin2020total,
  title={Total energy calculation for the metallic hcp phase of \textsc{Z}n in the bulk, layered, and quantum dot limits.},
  author={Olgu{\'\i}n, D},
  journal={Condensed Matter Physics},
  volume={23},
  number={3},
  year={2020}
}

@article{li2025electron,
  title={Electron transfer in catalysis: from fundamentals to strategies},
  author={Li, Xiaoning and Guan, Xinwei and Zhu, Lingfeng and Li, Hui and Yin, Xiaofeng and Sun, Shujie and Xu, Haimei and Fan, Yameng and Li, Peng and Hu, Long and others},
  journal={Chemical Society Reviews},
  year={2025},
  publisher={Royal Society of Chemistry}
}

@article{qi2023effect,
  title={Effect of isostructural phase transition on cycling stability of \textsc{Z}r\textsc{C}o-based alloys for hydrogen isotopes storage},
  author={Qi, Jiacheng and Liang, Zhaoqing and Xiao, Xuezhang and Yao, Zhendong and Zhou, Panpan and Li, Ruhong and Lv, Ling and Zhang, Xinyi and Kou, Huaqin and Huang, Xu and others},
  journal={Chemical Engineering Journal},
  volume={455},
  pages={140571},
  year={2023},
  publisher={Elsevier}
}

@article{yaghmaee2017thermodynamics,
  title={Thermodynamics modeling of cohesive energy of metallic nano-structured materials},
  author={Yaghmaee, Maziar Sahba and Baghbaderani, Hasan Ahmadian},
  journal={Materials \& Design},
  volume={114},
  pages={521--530},
  year={2017},
  publisher={Elsevier}
}

@article{kobayashi2018lattice,
  title={Lattice expansion and local lattice distortion in \textsc{N}b-and \textsc{L}a-doped \textsc{S}r\textsc{T}i\textsc{O}$_3$ single crystals investigated by x-ray diffraction and first-principles calculations},
  author={Kobayashi, Shunsuke and Ikuhara, Yuichi and Mizoguchi, Teruyasu},
  journal={Physical Review B},
  volume={98},
  number={13},
  pages={134114},
  year={2018},
  publisher={APS}
}

@article{chen2015phase,
  title={Phase stability, electronic, elastic and thermodynamic properties of \textsc{A}l-\textsc{RE} intermetallics in \textsc{M}g-\textsc{A}l-\textsc{RE} alloy: \textsc{A} first principles study},
  author={Chen, HL and Lin, L and Mao, PL and Liu, Z},
  journal={Journal of Magnesium and Alloys},
  volume={3},
  number={3},
  pages={197--202},
  year={2015},
  publisher={Elsevier}
}

@article{jacobs2002bulk,
  title={Bulk and surface properties of metallic aluminium: \textsc{DFT} simulations},
  author={Jacobs, PWM and Zhukovskii, Yu F and Mastrikov, Yu and Shunin, Yu N},
  journal={Comput. Model. New Technol},
  volume={6},
  pages={7--28},
  year={2002}
}

@article{nasui2025cold,
  title={A Cold Sintering Process for Manufacturing \textsc{Z}n Foams from Spherical Powders},
  author={Nasui, Mircea and Thalmaier, Gyorgy and Sechel, Niculina Argentina and Marinca, Traian Florin and R{\^\i}șteiu, Gabriela-Alexandra and Vida-Simiti, Ioan},
  journal={Applied Sciences},
  volume={15},
  number={22},
  pages={12179},
  year={2025},
  publisher={MDPI}
}

@article{masadeh2019total,
  title={Total-scattering pair-distribution function analysis of zinc from high-energy synchrotron data},
  author={Masadeh, Ahmad S and Shatnawi, Moneeb TM and Adawi, Ghosoun and Ren, Yang},
  journal={Modern Physics Letters B},
  volume={33},
  number={33},
  pages={1950410},
  year={2019},
  publisher={World Scientific}
}

@article{muralidhara2011electrodeposition,
  title={Electrodeposition of nanocrystalline Zinc on steel substrate from acid sulphate Bath and its corrosion study},
  author={Muralidhara, HB and Balasubramanyam, J and Naik, Y Arthoba and Kumar, K Yogesh and Hanumanthappa, H and Veena, MS},
  journal={J. Chem. Pharm. Res},
  volume={3},
  number={6},
  pages={433--449},
  year={2011}
}

@article{tamurejo2023electrodeposited,
  title={Electrodeposited zinc coatings for biomedical application: morphology, corrosion and biological behaviour},
  author={Tamurejo-Alonso, Purificaci{\'o}n and Gonz{\'a}lez-Mart{\'\i}n, Mar{\'\i}a Luisa and Pacha-Olivenza, Miguel {\'A}ngel},
  journal={Materials},
  volume={16},
  number={17},
  pages={5985},
  year={2023},
  publisher={MDPI}
}

@article{mamun2025improved,
  title={Improved photocatalytic activity of $\alpha$-\textsc{F}e$_2$\textsc{O}$_3$ by introducing \textsc{B}, \textsc{Y}, and \textsc{N}b dopants for solar-driven water splitting: a first-principles study},
  author={Mamun, Abdul Ahad and Talukder, Muhammad Anisuzzaman},
  journal={Materials Advances},
  volume={6},
  number={14},
  pages={4755-4767},
  year={2025},
  publisher={Royal Society of Chemistry}
}

@article{grimme2010consistent,
  title={A consistent and accurate \textit{ab initio} parametrization of density functional dispersion correction (\textsc{DFT-D}) for the 94 elements \textsc{H-P}u},
  author={Grimme, Stefan and Antony, Jens and Ehrlich, Stephan and Krieg, Helge},
  journal={The Journal of Chemical Physics},
  volume={132},
  number={15},
  year={2010},
  publisher={AIP Publishing}
}

@article{giannozzi2009quantum,
  title={QUANTUM ESPRESSO: a modular and open-source software project for quantumsimulations of materials},
  author={Giannozzi, Paolo and Baroni, Stefano and Bonini, Nicola and Calandra, Matteo and Car, Roberto and Cavazzoni, Carlo and Ceresoli, Davide and Chiarotti, Guido L and Cococcioni, Matteo and Dabo, Ismaila and others},
  journal={Journal of Physics: Condensed Matter},
  volume={21},
  number={39},
  pages={395502},
  year={2009},
  publisher={IOP Publishing}
}

@article{giannozzi2017advanced,
  title={Advanced capabilities for materials modelling with Quantum ESPRESSO},
  author={Giannozzi, Paolo and Andreussi, Oliviero and Brumme, Thomas and Bunau, Oana and Nardelli, M Buongiorno and Calandra, Matteo and Car, Roberto and Cavazzoni, Carlo and Ceresoli, Davide and Cococcioni, Matteo and others},
  journal={Journal of Physics: Condensed Matter},
  volume={29},
  number={46},
  pages={465901},
  year={2017},
  publisher={IOP Publishing}
}

@article{kresse1999ultrasoft,
  title={From ultrasoft pseudopotentials to the projector augmented-wave method},
  author={Kresse, Georg and Joubert, Daniel},
  journal={Physical Review B},
  volume={59},
  number={3},
  pages={1758},
  year={1999},
  publisher={APS}
}

@article{monkhorst1976special,
  title={Special points for Brillouin-zone integrations},
  author={Monkhorst, Hendrik J and Pack, James D},
  journal={Physical Review B},
  volume={13},
  number={12},
  pages={5188},
  year={1976},
  publisher={APS}
}

@article{lu2021first,
  title={First-principles study on the mechanical, thermal properties and hydrogen behavior of ternary V-Ni-M alloys},
  author={Lu, Yanli and Wang, Yi and Wang, Yifan and Gao, Meng and Chen, Yao and Chen, Zheng},
  journal={Journal of Materials Science \& Technology},
  volume={70},
  pages={83--90},
  year={2021},
  publisher={Elsevier}
}

@article{el2012corrosion,
  title={Corrosion study of zinc, nickel, and zinc-nickel alloys in alkaline solutions by Tafel plot and impedance techniques},
  author={El-Sayed, Abdel-Rahman and Mohran, Hossnia S and Abd El-Lateef, Hany M},
  journal={Metallurgical and Materials Transactions A},
  volume={43},
  number={2},
  pages={619--632},
  year={2012},
  publisher={Springer}
}

@article{heo2022modulation,
  title={Modulation of solvation structure and electrode work function by an ultrathin layer of polymer of intrinsic microporosity in zinc ion batteries},
  author={Heo, Jiyun and Hwang, Young-Eun and Doo, Gisu and Jung, Jinkwan and Shin, Kyungjae and Koh, Dong-Yeun and Kim, Hee-Tak},
  journal={Small},
  volume={18},
  number={25},
  pages={2201163},
  year={2022},
  publisher={Wiley Online Library}
}

@article{zeradjanin2017balanced,
  title={Balanced work function as a driver for facile hydrogen evolution reaction--comprehension and experimental assessment of interfacial catalytic descriptor},
  author={Zeradjanin, Aleksandar R and Vimalanandan, Ashokanand and Polymeros, George and Topalov, Angel A and Mayrhofer, Karl JJ and Rohwerder, Michael},
  journal={Physical Chemistry Chemical Physics},
  volume={19},
  number={26},
  pages={17019--17027},
  year={2017},
  publisher={Royal Society of Chemistry}
}

@article{chen2024boosting,
  title={Boosting oxygen evolution reaction by \textsc{F}e\textsc{N}i hydroxide-organic framework electrocatalyst toward alkaline water electrolyzer},
  author={Chen, Yuzhen and Li, Qiuhong and Lin, Yuxing and Liu, Jiao and Pan, Jing and Hu, Jingguo and Xu, Xiaoyong},
  journal={Nature Communications},
  volume={15},
  number={1},
  pages={7278},
  year={2024},
  publisher={Nature Publishing Group UK London}
}

@article{elsharkawy2025effect,
  title={Effect of \textsc{F}e/\textsc{N}i ratio on electrodeposition of \textsc{N}i-\textsc{F}e alloys and their bifunctional catalytic performance in hydrogen and oxygen evolution reactions},
  author={Elsharkawy, Safya and {\.Z}abi{\'n}ski, Piotr},
  journal={Journal of Power Sources},
  volume={660},
  pages={238516},
  year={2025},
  publisher={Elsevier}
}

@article{lim2020bimetallic,
  title={Bimetallic \textsc{N}i\textsc{F}e alloys as highly efficient electrocatalysts for the oxygen evolution reaction},
  author={Lim, Dongwook and Oh, Euntaek and Lim, Chaewon and Shim, Sang Eun and Baeck, Sung-Hyeon},
  journal={Catalysis Today},
  volume={352},
  pages={27--33},
  year={2020},
  publisher={Elsevier}
}

@article{zhu2020porous,
  title={Porous amorphous FeCo alloys as pre-catalysts for promoting the oxygen evolution reaction},
  author={Zhu, Wenjuan and Zhu, Guoxing and Yao, Chengli and Chen, Hu and Hu, Jing and Zhu, Yi and Liang, Wenfu},
  journal={Journal of Alloys and Compounds},
  volume={828},
  pages={154465},
  year={2020},
  publisher={Elsevier}
}

@article{shahzadi2025zn,
  title={\textsc{Z}n-\textsc{C}o nanoferrite electrocatalysts for enhanced hydrogen and oxygen generation},
  author={Shahzadi, Kiran and Sarfraz, Muhammad and Alomar, Muneerah and Mujtaba, MA and Bashir, Muhammad Nasir and Ali, Muhammad Mahmood and Ali, Faisal},
  journal={Results in Chemistry},
  pages={102392},
  year={2025},
  publisher={Elsevier}
}

@article{durmus2019influence,
  title={Influence of \textsc{A}l alloying on the electrochemical behavior of \textsc{Z}n electrodes for \textsc{Z}n--\textsc{A}ir batteries with neutral sodium chloride electrolyte},
  author={Durmus, Yasin Emre and Montiel Guerrero, Saul Said and Tempel, Hermann and Hausen, Florian and Kungl, Hans and Eichel, R{\"u}diger-A},
  journal={Frontiers in Chemistry},
  volume={7},
  pages={800},
  year={2019},
  publisher={Frontiers Media SA}
}

@article{kim2023electrocatalytic,
  title={Electrocatalytic properties of \textsc{C}o$_3$\textsc{O}$_4$ prepared on carbon fibers by thermal metal--organic deposition for the oxygen evolution reaction in alkaline water electrolysis},
  author={Kim, Myeong Gyu and Choi, Yun-Hyuk},
  journal={Nanomaterials},
  volume={13},
  number={6},
  pages={1021},
  year={2023},
  publisher={MDPI}
}

@article{yang2016fe3o4,
  title={\textsc{F}e$_3$\textsc{O}$_4$-decorated \textsc{C}o$_9$\textsc{S}$_8$ nanoparticles in situ grown on reduced graphene oxide: a new and efficient electrocatalyst for oxygen evolution reaction},
  author={Yang, Jing and Zhu, Guoxing and Liu, Yuanjun and Xia, Jiexiang and Ji, Zhenyuan and Shen, Xiaoping and Wu, Shikui},
  journal={Advanced Functional Materials},
  volume={26},
  number={26},
  pages={4712--4721},
  year={2016},
  publisher={Wiley Online Library}
}

@article{kim2019synergistic,
  title={Synergistic coupling derived cobalt oxide with nitrogenated holey two-dimensional matrix as an efficient bifunctional catalyst for metal--air batteries},
  author={Kim, Jeongwon and Gwon, Ohhun and Kwon, Ohhun and Mahmood, Javeed and Kim, Changmin and Yang, Yejin and Lee, Hansol and Lee, Jong Hoon and Jeong, Hu Young and Baek, Jong-Beom and others},
  journal={ACS Nano},
  volume={13},
  number={5},
  pages={5502--5512},
  year={2019},
  publisher={ACS Publications}
}

@article{garcia2025nife,
  title={\textsc{N}i\textsc{F}e and \textsc{N}i\textsc{C}o core-shell nanoparticles supported on graphene as efficient catalysts for oxygen evolution reaction},
  author={Garc{\'\i}a, Miriam L{\'o}pez and Mar{\'\i}a, Gonz{\'a}lez-Ingelmo and Usoltsev, Oleg and Oropeza, Freddy E and Timoshenko, Janis and Cuenya, Beatriz Roldan and Santamar{\'\i}a, Ricardo and Blanco, Clara and Rocha, Victoria G},
  journal={International Journal of Hydrogen Energy},
  volume={130},
  pages={313--323},
  year={2025},
  publisher={Elsevier}
}

@article{lankauf2020mnxco3,
  title={\textsc{M}n$_x$\textsc{C}o$_{3-x}$\textsc{O}$_4$ spinel oxides as efficient oxygen evolution reaction catalysts in alkaline media},
  author={Lankauf, Krystian and Cysewska, Karolina and Karczewski, Jakub and Mielewczyk-Gry{\'n}, A and G{\'o}rnicka, Karolina and Cempura, Grzegorz and Chen, Ming and Jasi{\'n}ski, Piotr and Molin, Sebastian},
  journal={International Journal of Hydrogen Energy},
  volume={45},
  number={29},
  pages={14867--14879},
  year={2020},
  publisher={Elsevier}
}

\end{document}